\documentclass[twocolumn,apl,amsmath,amssymb,showpacs]{revtex4-2}
\usepackage{epsf}
\usepackage{graphicx}
\usepackage{color}
\usepackage{soul}
\usepackage{gensymb}
\usepackage{sidecap}
\usepackage{amsmath}
\usepackage{multirow}
\usepackage[usestackEOL]{stackengine}
\begin{document}

\title{Multiple magnetic interactions and large inverse magnetocaloric effect in TbSi and TbSi$_{0.6}$Ge$_{0.4}$}

\author{Ajay Kumar} 
\email{ajay1@ameslab.gov}
\affiliation{Ames National Laboratory of the U.S. Department of Energy, Iowa State University, Ames, Iowa 50011, USA} 

\author{Prashant Singh}
\affiliation{Ames National Laboratory of the U.S. Department of Energy, Iowa State University, Ames, Iowa 50011, USA} 

\author{Andrew Doyle}
\affiliation{Ames National Laboratory of the U.S. Department of Energy, Iowa State University, Ames, Iowa 50011, USA}  

\author{Deborah L. Schlagel}
\affiliation{Ames National Laboratory of the U.S. Department of Energy, Iowa State University, Ames, Iowa 50011, USA} 

\author{Yaroslav Mudryk}
\affiliation{Ames National Laboratory of the U.S. Department of Energy, Iowa State University, Ames, Iowa 50011, USA} 

\date{\today}

\begin{abstract}

We present a comprehensive investigation of the electronic structure, magnetization, specific heat, and crystallography of TbSi (FeB structure type) and TbSi$_{0.6}$Ge$_{0.4}$ (CrB structure type) compounds. Both TbSi and TbSi$_{0.6}$Ge$_{0.4}$ exhibit two antiferromagnetic (AFM) transitions at T$_{\rm N1}\approx$ 58~K and 57~K, and T$_{\rm N2}\approx$ 36~K and 44~K, respectively, along with an onset of weak metamagnetic-like transition around 6~T between T$_{\rm N1}$ and T$_{\rm N2}$. High-resolution specific heat (C$_{\rm P}$) measurements show the second- and first-order nature of the magnetic transition at T$_{\rm N1}$ and T$_{\rm N2}$, respectively, for both samples. However, in the case of TbSi, the low-temperature (LT) AFM to high-temperature (HT) AFM transition takes place via an additional AFM phase at the intermediate temperature (IT), where both LT to IT AFM and IT to HT AFM phase transitions exhibit a first-order nature. Both TbSi and TbSi$_{0.6}$Ge$_{0.4}$ manifest significant magnetic entropy changes ($\Delta S_{\rm M}$) of 9.6 and 11.6~J/kg-K, respectively, for $\Delta \mu_0H$=7~T, at T$_{\rm N2}$. The HT AFM phase of TbSi$_{0.6}$Ge$_{0.4}$ is found to be more susceptible to the external magnetic field, causing a significant broadening in the peaks of $\Delta S_{\rm M}$ curves at higher magnetic fields. Temperature and field-dependent specific heat data have been utilized to construct the complex H-T phase diagram of these compounds. Furthermore, temperature-dependent x-ray diffraction measurements demonstrate substantial magnetostriction and anisotropic thermal expansion of the unit cell in both samples.

\end{abstract}

\maketitle
 
\section{\noindent ~Introduction}

Rare earth (RE)-Si-Ge compounds with varying stoichiometry have been a subject of considerable interest within the scientific community for several decades due to their intriguing properties, such as giant magnetocaloric effect (GMCE), colossal magnetoresistance (CMR), metal-insulator transition (MIT), Griffiths phase, kinetically arrested magnetostructural transitions, and anomalous Hall effect, among others \cite{Pecharsky_PRL_97, Liu_PRB_03, Zou_PRB_09, Mohapatra_PRB_09, Jammalamadaka_PRBR_09, Magen_PRL_06, Tseng_PRB_11, Pereira_PRB_10, Teizer_PRL_2000, Stankiewicz_PRB_2000,  Lyu_PRB_20}. In this family of materials, the RE$_5$(Si$_x$Ge$_{1-x}$)$_4$ compounds have garnered extensive attention, especially after the discovery of GMCE near room temperature in Gd$_5$(Si$_2$Ge$_2$), a phenomenon associated with the peculiar first-order magnetostructural phase transition at $\sim$276~K \cite{Pecharsky_PRL_97, Pereira_PRB_10, Choe_PRL_2000, Mozharivskyj_JACS_05}. The exchange interactions between the RE ions in these compounds are indirect RKKY-type, mediated by conduction electrons. The sign of these interactions is determined by the distance between the rare earth ions, making them susceptible to external perturbations such as temperature, hydrostatic pressure, chemical pressure (doping), and magnetic field \cite{Tseng_PRB_11, Mohapatra_PRB_09, Mudryk_PRB_05, Tung_PRB_05}. A strong coupling between the crystal structure and magnetic properties have been observed in Gd$_5$(Si$_x$Ge$_{1-x}$)$_4$ pseudo-binary system. These compounds with $0 < x \leq 0.2$ on heating show a first-order transition from a low-temperature ferromagnetic (FM) state having a Gd$_5$Si$_4$ [O(I)]-type crystal structure to an intermediate antiferromagnetic (AFM) state with a Sm$_5$Ge$_4$ [O(II)]-type structure, and then to a high-temperature paramagnetic (PM) phase having O(II)-type structure. In these Ge-rich compounds the Curie temperature (T$_{\rm C}$) of FM-AFM transformation increases and the N\'eel temperature (T$_{\rm N}$) remains nearly unchanged with increasing $x$ \cite{Bunzli_book_05, Morellon_PRB_2000, Pecharsky_APL_97}. For $0.2 < x \leq 0.38$ there is a two-phase mixture of O(II) and monoclinic structures at room temperature, and for $0.38 < x \leq 0.5$, the system exhibits a first-order transformation from the same low-temperature O(I)-type FM phase to a high-temperature PM phase with a monoclinic crystal structure. Finally, Gd$_5$(Si$_x$Ge$_{1-x}$)$_4$ shows a second order transition from FM  to PM state with the same O(I)-type crystal structure in both the magnetic phases for $0.5 < x \leq 1$  \cite{Bunzli_book_05, Morellon_PRB_2000, Pecharsky_APL_97}. \par

The markedly different electronic structure of Tb$^{3+}$ ($4f^8$; S=L=3, J=6), as compared to Gd$^{3+}$ ($4f^7$; J=S=7/2, L=0) ions having isotropic spin-only magnetic moments, results in the strong directional character in Tb-based compounds due to L$\neq$0 \cite{Zou_PRB_09, Zou_PRB_07, Liu_PRB_03, Zou_PRB_09, Mohapatra_PRB_09, Jammalamadaka_PRBR_09, Magen_PRL_06, Ritter_PRB_02, Zou_PRB_08, Huang_SR_22}. Therefore, Gd is often substituted with Tb ions due to their similar ionic radii and valence state (both virtually trivalent) \cite{Magen_PRL_06, Morellon_PRB_03, Ritter_PRB_02} to study the effects of single-ion anisotropy on physical behavior of materials.  For instance, Tb$_5$Si$_{2.2}$Ge$_{1.8}$ shows a large anisotropic temperature and field dependent magnetization, and hence magnetocaloric effect, where a GMCE is observed along the $a$(broad)- and $c$ (sharp)-axes, but not along the $b$-axis \cite{Zou_PRB_07, Zou_PRB_08}. An anisotropic electrical resistivity and magnetoresistance are also observed in the Tb$_5$Si$_{2.2}$Ge$_{1.8}$ with a CMR of 160\% only obtained along the $a$-axis, whereas $b$- and $c$- axes display a significantly smaller MR of around 8\% and 5\%, respectively \cite{Zou_PRB_09}. Similarly, the anisotropic magnetic exchange interactions with positive Weiss temperature ($\theta_{\rm P}$) (FM interactions) along $a$- and $c$-axes, and negative (AFM coupling) along $b$-axis have been observed in Tb$_5$Ge$_4$, unlike Gd$_5$Ge$_4$, where $\theta_{\rm P}$ remains positive along all crystallographic directions \cite{Tian_PRB_09, Ouyang_PRB_06}.   This change in magnetic interactions in Tb$_5$(Si$_x$Ge$_{1-x}$)$_4$, despite similar RE-RE distances as compared to Gd$_5$(Si$_x$Ge$_{1-x}$)$_4$ is attributed to the large orbital moment and hence directional character of Tb$^{3+}$ ions. Complex low-temperature phase diagrams have also been observed in Sm$_5$Si$_x$Ge$_{4-x}$ \cite{Ahn_PRB_07}, Ho$_5$(Si$_x$Ge$_{1-x}$)$_4$ \cite{Ritter_PRB_09}, Yb$_5$(Si$_x$Ge$_{1-x}$)$_4$ \cite{Voyer_PRB_06}, etc., by virtue of change in ionic radii, and electronic configurations of localized 4$f$ and itinerant conduction electrons, along the lanthanide series. \par

In this context, the systematic investigation of pseudo-binary compounds RE-Si$_x$Ge$_{1-x}$ adds a crucial piece of knowledge to our understanding and control of the fundamental and functional properties of rare earth intermetallic compounds \cite{Tung_PRB_05, Das_PRB_14, Das_JPCM_12, Engkagul_PRB_87}. These compounds crystallize in either CrB-type $Cmcm$ or closely related FeB-type $Pnma$ structure and display complex composition-temperature ($x$-T) phase diagrams \cite{Papamantellos_JMMM_86, Papamantellos_JMMM_88, Papamantellos_JMMM_93, Papamantellos_JSSC_87, Papamantellos_JMMM_11, Thuery_JMMM_93}. In HoSi$_x$Ge$_{1-x}$, the $x =$1 sample shows both CrB (35.5\%)- and  FeB (64.5\%)- type structures at room temperature, whereas other members adopt CrB-type only \cite{Papamantellos_JSSC_87, Papamantellos_JMMM_11}. The $x>0.2$ samples display two distinct AFM structures, where the low-temperature (LT) lock-in phase with wave vector $q=(1/2, 0, 1/2)$ undergoes a first-order phase transition (FOPT) into a high-temperature (HT) incommensurate structure associated with  $q=(q_x$, 0, $q_z)$ at $T_{\rm ic}$, and finally into PM state via a second-order phase transformation (SOPT) at $T_{\rm N}$ \cite{Papamantellos_JSSC_87}. Both $T_{\rm ic}$ and $T_{\rm N}$ increase with $x$ \cite{Papamantellos_JSSC_87}. On the other hand, the magnetic structure of $x \leq 0.2$ samples  remains incommensurate down to 1.2~K \cite{Papamantellos_JSSC_87}. The ErSi$_x$Ge$_{1-x}$ shows even more complex $x$-T phase diagram with no Lifshitz point, and two commensurate and two incommensurate magnetic structures, which vary with both $x$ and T  \cite{Thuery_JMMM_93}. The application of magnetic field (H) can provide an additional degree of freedom to manipulate these $x$-T phase diagrams.  For instance, a strong coupling between localized 4$f$ moments and conduction electrons lead to anomalous magnetic field-induced phase transitions in GdSi at 2.5, 3.8, and 19.5 T \cite{Tung_PRB_05}. Feng et al. later reported the comprehensive H-T phase diagram of GdSi, revealing four different types of AFM interactions in the sample, the strengths and directions of which depend strongly on lattice directions \cite{Feng_PRB_13}. Several experimental and theoretical investigations have aimed to understand the metal-insulator transition in amorphous Gd-Si systems \cite{Bokacheva_PRB_04, Majumdar_PRL_03, Teizer_PRL_2000}. Moreover, large magnetic anisotropy and GMCE have been observed in single crystals of PrSi and PrGe \cite{Das_PRB_14, Das_JPCM_12}. \par

P. Schobinger-Papamantellos $et$ $al.$ reported the crystal and magnetic structure of TbSi$_x$Ge$_{1-x}$ for $0 \leq x \leq 1$ using temperature-dependent neutron diffraction measurements \cite{Papamantellos_JMMM_86, Papamantellos_JMMM_88, Papamantellos_JMMM_93}. All samples, except $x = 1$, crystallize in the orthorhombic CrB-type structure. The $0 \leq x < 0.6$ samples exhibit LT planar AFM interactions (in $ac$ plane) with wave vector $q =$ (1/2, 0, 1/2), which undergo a HT uniaxial AFM (along $c$ axis) having $q =$ (0, 0, 1/2) (at T$_{ic}$) via first-order transition, and finally experience a second-order AFM to PM transition (at T$_{\rm N}$) \cite{Papamantellos_JMMM_86, Papamantellos_JMMM_88}.  The $0.6 > x > 1$ samples, on the other hand, show a transition from LT AFM with $q =$ (1/2, 0, 1/2) to a sinusoidally modulated incommensurate AFM  structure having q= (q$_x$, 0, q$_z$) at T$_{ic}$ \cite{Papamantellos_JMMM_88}. More importantly, at the critical doping of $x =$0.6 (TbSi$_{0.6}$Ge$_{0.4}$),  both commensurate and incommensurate magnetic structures coexist in equal proportion in the HT phase region \cite{Papamantellos_JMMM_88}. Unlike HoSi$_x$Ge$_{1-x}$, the   T$_{\rm N}$ in the case of TbSi$_x$Ge$_{1-x}$ remains almost invariant; however, T$_{ic}$ increases with $x$ \cite{Papamantellos_JSSC_87, Papamantellos_JMMM_88}. Further, TbSi ($x = 1$) crystallizes in the FeB-type structure and displays even more complex magnetic interactions. The low-temperature phase (below 36 K) is planar AFM (in the $ab$ plane) with $q =$ (0, 1/2, 0), which exhibits a first-order transition into a three-dimensionally modulated incommensurate AFM phase with $q =$ (q$_x$, q$_y$, q$_z$) \cite{Papamantellos_JMMM_93}. Interestingly, the incommensuration persists predominantly up to 42 K only, and above this temperature, it tends to commensurate with q $\approx$ (0, 1/2, 1/8) up to T$_N$ = 57 K, followed by a second-order AFM to PM phase transition \cite{Papamantellos_JMMM_93}. Therefore, it is vital to study the the nature and behavior of this additional incommensurated magnetic structure in case of the TbSi in this narrow temperature range. Further, the detailed study of these multiple magnetic interactions and their evolution with an external magnetic field is missing together with the theoretical insights into the various unresolved magnetic phenomena in the RE-Si-Ge systems. In particular, the response of the observed first-order transition to applied magnetic field is important for understanding the potential for magnetocaloric effect and other magnetoresponsive phenomena in the underlined compounds \cite{Remya_JRE_23, Remya_JalCom_24, Bykov_JalCom_24}. \par

In this manuscript, we theoretically analyzed the structural stability and experimentally performed detailed magnetization, specific heat, and temperature-dependent x-ray diffraction (XRD) measurements on the two most intriguing members of this series, namely the FeB-type TbSi and the CrB-type TbSi$_{0.6}$Ge$_{0.4}$ samples. We observe a significant broadening in the magnetic entropy change ($\Delta S_{\rm M}$) curves in the case of TbSi$_{0.6}$Ge$_{0.4}$ at higher magnetic fields compared to TbSi. The high-resolution specific heat measurements are used to uncover this phenomenon, which indicates the evolution of additional field-induced AFM interactions in the HT AFM phase region of this sample. Furthermore, we unravel an additional first-order magnetic transition in the vicinity of the LT AFM phase of TbSi. Both samples exhibit finite magnetostriction at both LT and HT AFM transitions and demonstrate anisotropic thermal expansion of the unit cell in the magnetically ordered as well as PM state. We construct complex H-T phase diagrams for TbSi and TbSi$_{0.6}$Ge$_{0.4}$ using the temperature- and field-dependent specific heat data.

\section{\noindent ~Methodology}

\subsection{Theory}

We employed density-functional theory (DFT) as implemented within Vienna ab initio simulation package (VASP) \cite{Kresse_PRB_99} for electronic-structure calculation based on the Perdew, Burke, and Ernzerhof (PBE) exchange-correlation functionals \cite{Perdew_PRL_96}. The energy and Hellmann-Feynman force convergence criteria for full relaxation (volume and atomic coordinates) are 10$^{-7}$ eV and 10$^{-6}$ eV/\AA, respectively. A kinetic-energy cutoff of 520 eV was used for the plane-wave basis set. Spin-polarized calculations were performed due to the presence of magnetic elements. To enforce the localization of the $f$-electrons, the effective onsite Coulomb (U) and exchange (J) interaction parameters are set to be 6.00 and 0.90 eV, respectively \cite{Dudarev_PRB_98}. Monkhorst-Pack \cite{Monkhorst_PRB_76} k-mesh grids of 3×5×4 and 2×1×2 was used to sample the Brillouin zone in $Pnma$-TbSi and $Cmcm$-TbSi$_{0.6}$Ge$_{0.4}$ phases, respectively, during relaxation, while the mesh-size was doubled to accurately determine formation enthalpy and electronic-structure properties. The partially ordered TbSi$_{1-x}$Ge$_x$ ($x =$0--1) structures were created using Alloy Theoretic Automated Toolkit (ATAT) \cite{Walle_CH_02} to assess their thermodynamic (formation enthalpy) and structural (phonon) stability. The formation energy (E$_{\rm form}$=E$_{\rm tot}$-$\sum_i$n$_i$E$_i$, i=Tb,Ge,Si) was used to determine the phase stability (at 0~K) of TbSi$_{1-x}$Ge$_x$ ($x =$0--1) intermetallics. The E$_{\rm tot}$ is total energy of disorder alloys while \{n$_i$, E$_i$\} are number of atoms of type “i” in their ground state unit cell, i.e., \{Tb, Ge, Si\}. The phonon spectra and density of states are calculated using the finite displacement method as implemented in Phonopy software. The density-functional perturbation theory (DFPT) was used to construct force constant matrix needed to calculate phonon dispersion \cite{Togo_SM_15}. \par

\subsection{Experimental}

The TbSi and TbSi$_{0.6}$Ge$_{0.4}$ samples were prepared by arc melting stoichiometric amounts of pure Tb, Si, and Ge elements in an argon (Ar) atmosphere. The terbium metal was provided by the Materials Preparation Center of Ames National Laboratory and had an overall purity better than 99.9 \% with respect to all elements, silicon was supplied by Alfa Aesar (99.9995 \%), and germanium by Meldform Metals ($\sim$99.99 \%). Prior to the sample preparation, a Zr getter was melted to minimize the presence of residual or impurity gases in the chamber. Each sample was flipped and remelted 4-5 times to ensure the homogeneity. The weight loss during the melting process was found to be less than 1 wt. \% for both samples. While the TbSi was found to be single phase after melting, the TbSi$_{0.6}$Ge$_{0.4}$ sample required additional heat-treatment and was post annealed at 900 \degree C for 7 days in an highly pure helium-filled quartz tube.  Room temperature x-ray powder diffraction (XRD) patterns were recorded for both samples using a PANalytical X'Pert Pro diffractometer with Cu-K$_{\alpha}$ ($\lambda$ =1.5406 Å) radiation in the Bragg-Brentano geometry. Low-temperature XRD data were collected in the temperature range of 10-300 K using a Rigaku TTRAX powder diffractometer equipped with a rotating anode Mo K$_{\alpha}$ source and a continuous helium flow cryostat \cite{Holm_RSI_04}. The data were recorded in cooling mode after stabilizing the sample at each temperature. The Rietveld refinement of the XRD data was performed using the FullProf suite, employing a pseudo Voigt peak shape and linear interpolation between background points. The refinement process started with the 10~K data, followed by the Rietveld refinement of the high-temperature data, where the lattice parameters, overall B-factors, and atomic positions were the only varying parameters \cite{Kumar_JPCL_22}. Temperature and magnetic field-dependent magnetization measurements were conducted using a superconducting quantum interference device (SQUID) system from Quantum Design, USA (model MPMS XL-7). The temperature dependence of magnetization was measured using the following protocols: on warming in applied magnetic field after zero-field cooling (ZFC), on cooling in applied magnetic field (FCC), and on warming in applied magnetic field after cooling in the same applied field (FCW). To record the field dependent magnetization isotherms at different temperatures, we reduce the magnetic field from 70~kOe to 500~Oe in linear mode and then from 500~Oe to 0~Oe in the oscillatory mode before moving to next temperature (in heating mode) to avoid any remanence in the samples, keeping their low coercivity ($\sim$10~Oe) in mind. The temperature and field dependent specific heat measurements were performed using a physical property measurement system (PPMS) from Quantum Design, Inc, employing both standard relaxation technique as well as long heat pulse method. Further details about the measurement protocols are provided along with the discussion of the specific heat data.

\section{\noindent ~Results and discussion}

\subsection{Structural stability}

\begin{figure} 
\includegraphics[width=3.5in]{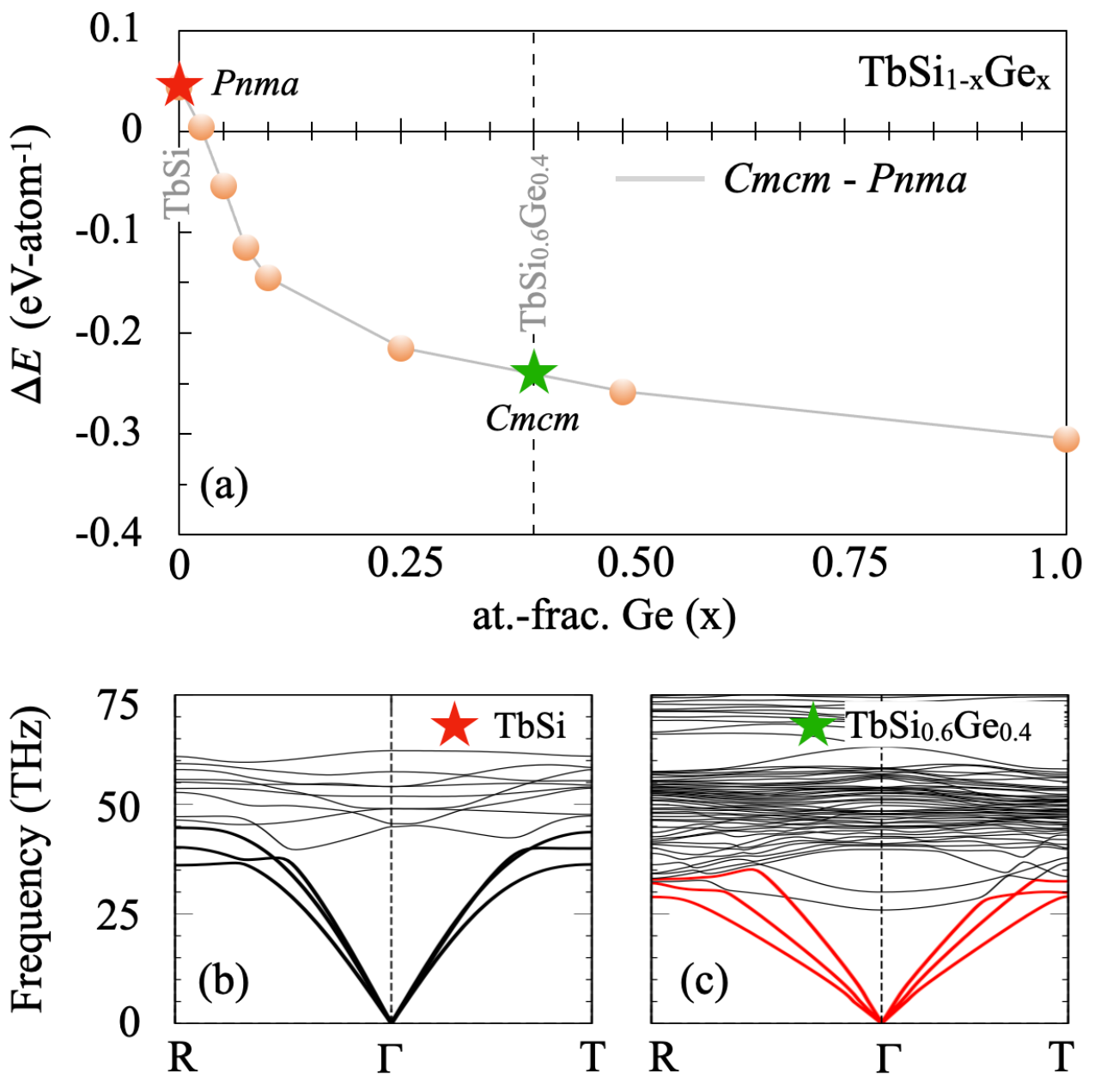}
\caption {(a) The formation energy difference ($\Delta$E$_{\rm form}$) between $Pnma$ and $Cmcm$ crystal structures for partially disordered TbSi$_{1-x}$Ge$_x$ ($x =$0--1) compounds calculated using DFT. (b, c) Phonon dispersion of TbSi and TbSi$_{0.6}$Ge$_{0.4}$ in  $Pnma$ and $Cmcm$ space group, respectively.}
\label{Fig1_DFT}
\end{figure}

In order to understand the structural stability of pseudo-binary TbSi$_{1-x}$Ge$_x$ compounds, we plot the formation energy difference ($\Delta$E$_{\rm form}$) between the $Pnma$ (FeB-type) and $Cmcm$ (CrB-type) orthorhombic phases in Fig. \ref{Fig1_DFT}(a) for the $x =$0--1. The calculated E$_{\rm form}$ of the $x$=0 is -0.78 and -0.75 eV/atom, and  for $x =$1 is -0.57 and -0.88 eV/atom in the $Pnma$ and $Cmcm$ phases, respectively, which indicate that TbSi ($x =$0) and TbGe ($x =$1) compounds stabilize in the $Pnma$ and $Cmcm$ structures, respectively, consistent with the reported neutron diffraction (ND) data \cite{Papamantellos_JMMM_86, Papamantellos_JMMM_88, Papamantellos_JMMM_93}. The ND measurements provide a lower limit of $x= $0.1 (10\% Ge), sufficient to transform the TbSi from the $Pnma$ to $Cmcm$ structure \cite{Papamantellos_JMMM_88}. Therefore, in order to further pinpoint the critical limit of Ge required for this structural transformation in TbSi$_{1-x}$Ge$_x$, we calculate the $\Delta$E$_{\rm form}$ in a step of 2.5\% in the critical regime ($x =$0--0.1), as shown in Fig. \ref{Fig1_DFT} (a). Interestingly, the $\Delta$E$_{\rm form}$ shows a crossover from $Pnma$ to $Cmcm$ orthorhombic phase at 2.5 at.\% Ge, i.e., the $Pnma$ phase is stable only from 0 to 2.5 at. \% Ge, while $Cmcm$ phase is energetically more favourable for Ge $>$2.5 at.\%. Now, we will focus on the FeB-type ($Pnma$) TbSi and CrB-type ($Cmcm$) TbSi$_{0.6}$Ge$_{0.4}$ samples, explored in this study, to understand why Ge-doped TbSi$_{0.6}$Ge$_{0.4}$ adopts the $Cmcm$ structure instead of the parent crystal structure of binary TbSi. We hypothesize that the origin of this change lies in local structural features such as atomic distortion or displacement. Before delving deeper into the discussion, for the clarity of readers, we want to emphasize that terbium, being in an open $f$-shell structure with a $+3$ oxidation state (i.e., $4f^{8}$ electronic state), has its magnetic nature accounted for by placing $f$-states into the valence in both thermodynamic and electronic structure evaluations \cite{Vogel_2019}. To study the local distortion, the TbSi$_{0.6}$Ge$_{0.4}$ composition was fully relaxed in $Cmcm$ and $Pnma$ structures using DFT to obtain the optimized atomic coordinates. The $Cmcm$ phase was kept at a fixed experimental volume. For the $Pnma$-TbSi$_{0.6}$Ge$_{0.4}$ phase created using the $Pnma$-TbSi crystal structure, we performed full lattice relaxation to obtain the optimal volume and coordinates. Using the experimental crystal structure as a reference, the average vector atomic displacement in the $Cmcm$ and $Pnma$ phases of TbSi$_{0.6}$Ge$_{0.4}$ was calculated as 0.314 and 0.228, respectively. The method to estimate atomic displacements in complex lattices has been discussed in Ref. \cite{Singh_AM_23}. The increased vector atomic displacement in the $Cmcm$ phase suggests higher charge activity, i.e., higher distortion, improves the phase stability, as the $Cmcm$ phase has higher crystal symmetry compared to $Pnma$. The higher local distortion in the case of TbSi$_{0.6}$Ge$_{0.4}$ agrees with experimental structural analysis discussed below, and is consistent with a recent study where the degree of distortion was found to be correlated with both bond strength and phase stability in disordered materials \cite{Singh_AM_23}.\par

The dynamical stability of TbSi and TbSi$_{0.6}$Ge$_{0.4}$ compounds is further confirmed via phonon dispersion curves, as shown in Figs. \ref{Fig1_DFT}(b, c), respectively. The dispersion shows no obvious imaginary phonon modes for the two experimentally synthesized phases. The low-frequency contribution in TbSi comes from mixed Tb/Si atoms, while the Ge-doped case shows phonon softening. The low, mid, and high-frequency contributions in TbSi$_{0.6}$Ge$_{0.4}$ come from Tb, Ge, and Si, respectively [see Fig. \ref{Fig1_DFT}(c)]. This suggests that Tb atoms make the main contribution to the low-energy phonons, which brings electron-phonon coupling at low temperature and drives interesting magnetic characteristics. Our argument is also consistent with previous reports, where adding a high Z element was found to soften the phonons \cite{Kargar_2018, Delaire_2011}. In Fig. \ref{Fig1_DFT}(c), the Ge doping changes the Tb-Si interaction, which weakly softens the phonon modes as clearly visible through the reduced acoustic mode frequency in the 25 to 50 THz range. The Ge doping also creates Ge-Si interactions that lead to the spread of overall optical phonon modes, as seen by high-energy phonons, which arise mainly from increased Ge/Si interaction. This also indicates that the phonon free energies of TbSi$_{0.6}$Ge$_{0.4}$ are smaller than those of TbSi due to the monotonicity of free energy with respect to the phonon frequency, which leads to phonon softening and stabilizes this intermetallic compound in the $Cmcm$ phase. To further affirm our thermodynamic stability of the $Cmcm$ phase in TbSi$_{0.6}$Ge$_{0.4}$, we present the phonon spectra of $Pnma$-TbSi$_{0.6}$Ge$_{0.4}$ in Fig. 1 of \cite{SI}, which shows large imaginary phonons at $\Gamma$ points, indicative of structural instability. The origin of this energy instability of the $Pnma$ structure for TbSi$_{0.6}$Ge$_{0.4}$ lies in the local charge rearrangement at the Tb-site. We found that Tb $f$-states show significant charge gain (-0.204e), resulting in more $f$-states at the Fermi level, and hence local charge instability. On the other hand, Tb $d$-states, which play a key role in phase stability through significant $d$-$f$ exchange in these compounds, weaken due to noteworthy charge loss from Tb-5$d$ (-0.501e) on Ge substitution when compared to the $Cmcm$ phase of TbSi$_{0.6}$Ge$_{0.4}$.

\subsection{Room-temperature crystallography}

\begin{figure} 
\includegraphics[width=3.5in]{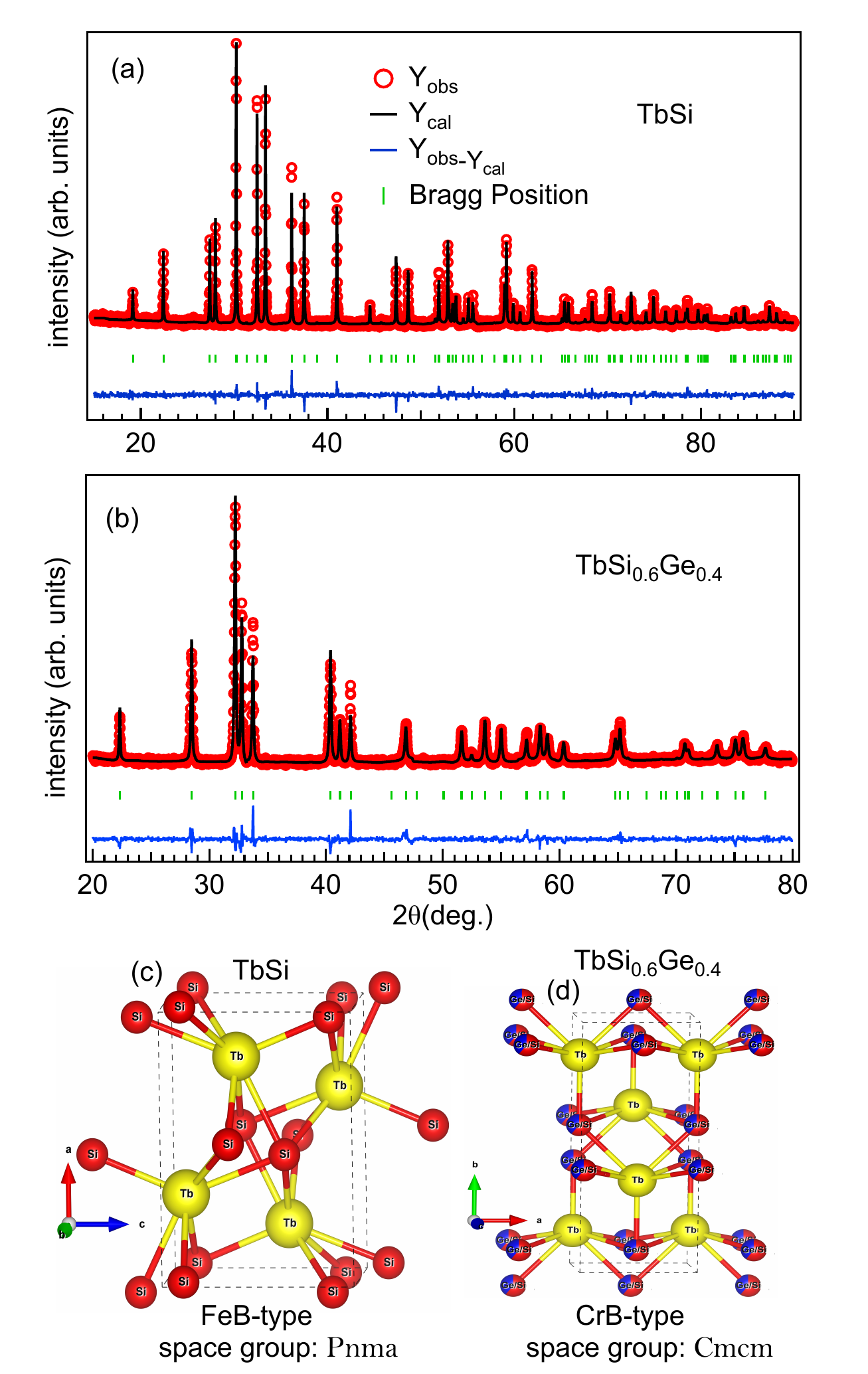}
\caption {The Rietveld refinement of room temperature XRD data of (a) TbSi and (b) TbSi$_{0.6}$Ge$_{0.4}$ samples measured using Cu K$_{\alpha}$ radiation, where the observed data points, simulated curves, Bragg positions, and the difference in observed and calculated data are shown by open red circles, a continuous black line, vertical green bars, and a continuous blue line, respectively. (c, d) The 3D illustration of the FeB-type ($Pnma$ space group) and CrB-type ($Cmcm$ space group) crystal structures of TbSi and TbSi$_{0.6}$Ge$_{0.4}$, respectively.} 
\label{Fig2_RTXRD}
\end{figure}

The results of the Rietveld refinement of the room temperature x-ray powder diffraction patterns of TbSi and TbSi$_{0.6}$Ge$_{0.4}$ samples are presented in Figs. \ref{Fig2_RTXRD}(a) and (b), respectively. The data confirm that  substitution of Ge at Si site changes the crystal structure from FeB-type ($Pnma$ space group) in TbSi to CrB-type ($Cmcm$ space group) in TbSi$_{0.6}$Ge$_{0.4}$, consistent with our theoretical simulations (discussed above) and   Refs. \cite{Papamantellos_JMMM_86, Papamantellos_JMMM_88, Papamantellos_JMMM_93}.  The extracted room temperature lattice parameters are $a$= 7.918 \AA $\space$ and 4.288 \AA, $b$= 3.835 \AA $\space$ and 10.615 \AA, and $c$= 5.703 \AA $\space$ and 3.875 \AA $\space$ for  TbSi and TbSi$_{0.6}$Ge$_{0.4}$, respectively. The three dimensional crystal structure plots of the TbSi and TbSi$_{0.6}$Ge$_{0.4}$ compounds are presented in Figs. \ref{Fig2_RTXRD} (c) and (d), respectively, with their common axis of symmetry, i.e., the $b$-axis of TbSi and the $c$-axis of TbSi$_{0.6}$Ge$_{0.4}$, shown out of the plane of the paper. Each unit cell [represented by the dashed cuboids in Figs. \ref{Fig2_RTXRD} (c) and (d)] contains the eight atoms (four formula units) in both the cases. In the TbSi (FeB-type structure), both Tb and Si atoms occupy 4$c$ ($x$, 1/4, $z$) position, where $x$=0.180 and 0.014, and $z$= 0.616 and 0.132, respectively; whereas in the TbSi$_{0.6}$Ge$_{0.4}$ (CrB-type structure), all atoms occupy the 4$c$ (0, $y$, 1/4) position with $y$=0.141 and 0.418 for Tb and Si/Ge, respectively, at room temperature.  

\subsection{Magnetic properties and magnetocaloric effect}

\begin{figure}  
\centering
\includegraphics[width=3.5in]{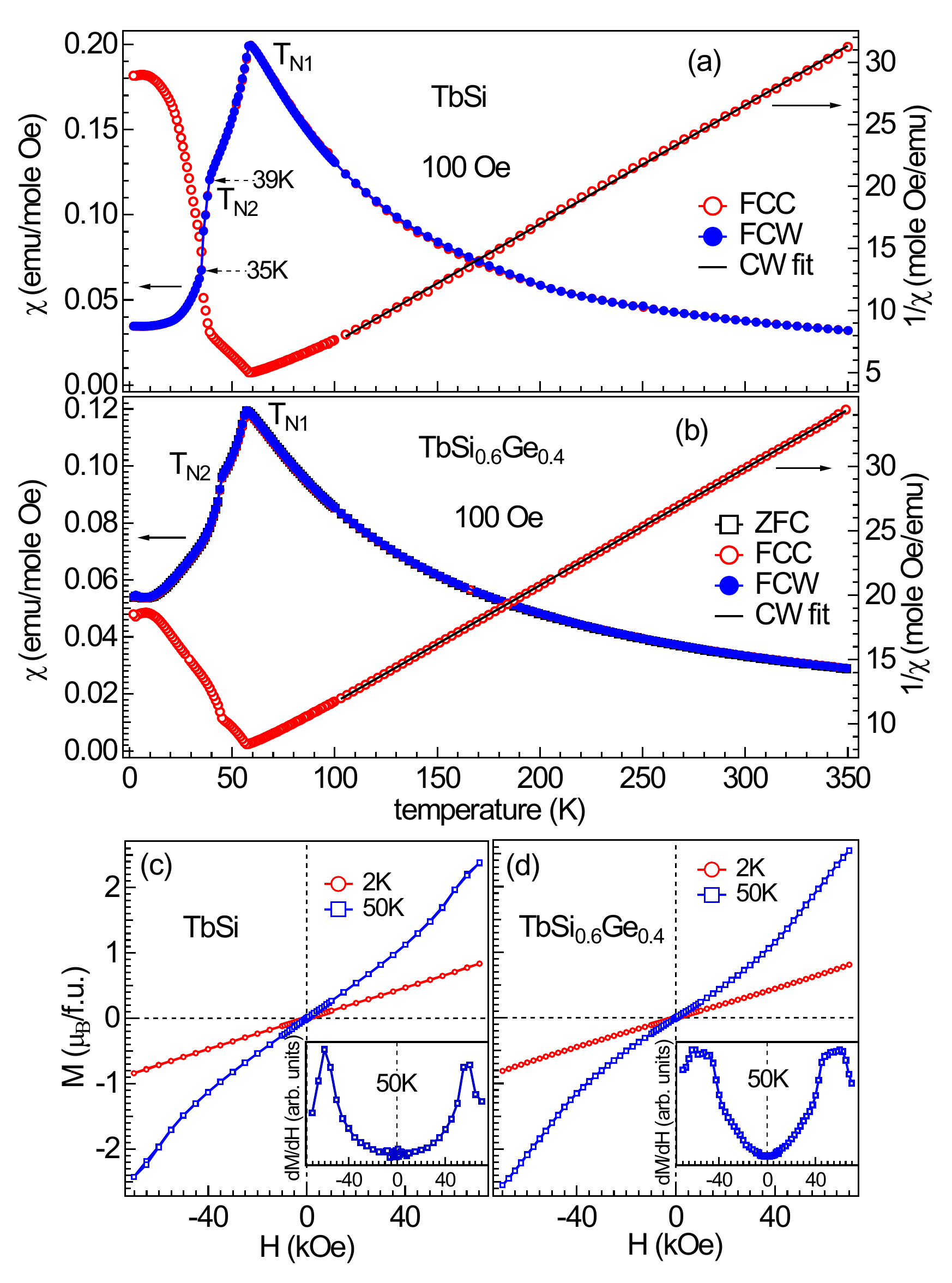}
\caption {Temperature-dependent direct (left axis) and inverse (right axis) dc magnetic susceptibility for (a) TbSi and (b) TbSi$_{0.6}$Ge$_{0.4}$ compounds measured at 100 Oe in ZFC, FCW, and FCC modes. Solid black lines represent the fit of the FCC curves from 105 to 350~K using the Curie-Weiss law. Dashed arrows in (a) represent the abrupt enhancement in the susceptibility of TbSi between 35 to 39~K. The field-dependent magnetization curves of (c) TbSi and (d) TbSi$_{0.6}$Ge$_{0.4}$ samples, recorded at 2~K and 50~K, where insets represent the dM/dH versus H plot for one -70~kOe to +70~kOe M-H cycle at 50~K.}

\label{Fig3_MT_MH}
\end{figure}

In order to understand the influence of Ge substitution on the magnetic properties, temperature-dependent magnetization data have been recorded between 2--350~K for both TbSi and TbSi$_{0.6}$Ge$_{0.4}$ in ZFC, FCW, and FCC modes at 100~Oe, as presented in Figs. \ref{Fig3_MT_MH} (a) and (b), respectively. Both samples show two well-resolved antiferromagnetic transitions around 58$\pm$1~K (called T$_{\rm N1}$) and 39$\pm$1~K (called T$_{\rm N2}$) for TbSi, and 57$\pm$1~K and 45$\pm$1~K for TbSi$_{0.6}$Ge$_{0.4}$ [see Figs. \ref{Fig3_MT_MH}(a, b)]. It is interesting to note that an increase in the Si concentration in TbSi$_x$Ge$_{1-x}$ samples was found to stabilize the low-temperature (LT) planar AFM (increase in T$_{\rm N2}$) for $x \leqslant$0.9, having a CrB-type crystal structure \cite{Papamantellos_JMMM_88}. However, a reduction in T$_{\rm N2}$ in the case of TbSi having FeB-type structure, as compared to TbSi$_{0.6}$Ge$_{0.4}$, indicates that the transformation in crystal structure also influences the strength of LT AFM in these samples. Although, no significant change in the strength of the high-temperature (HT) AFM (T$_{\rm N1}$) is found between TbSi and TbSi$_{0.6}$Ge$_{0.4}$, analogous to the other members of the TbSi$_x$Ge$_{1-x}$ series \cite{Papamantellos_JMMM_88}. Moreover, we observe an abrupt enhancement in the magnetic moment in the case of TbSi between 35-39~K (in the vicinity of T$_{\rm N2}$), as indicated by the dashed arrows in Fig. \ref{Fig3_MT_MH}(a), suggesting a possible change in the magnetic structure of TbSi in this narrow temperature range. The neutron diffraction measurements also show a drastic change in the wave vector of TbSi from q=(0, 1/2, 0) to a three-dimensionally modulated AFM structure with q=(q$_x$, q$_y$, q$_z$) in the 36-39~K temperature range \cite{Papamantellos_JMMM_93}. We will discuss this point in more detail while discussing the high-resolution specific heat data in the next section. Furthermore, the ND measurements reported the first-order nature of the LT to HT AFM transition in both of these samples \cite{Papamantellos_JMMM_88, Papamantellos_JMMM_93}. However, no notable hysteresis has been observed in the magnetization data recorded in the heating and cooling modes (ZFC/FCW and FCC) in the entire temperature range for both samples. A larger temperature step (1~K) compared to the thermal hysteresis ($<$0.5~K, discussed below) along with the steep increase in magnetization at T$_{\rm N2}$ may be responsible for this. The inverse magnetic susceptibility of the FCC curves is plotted on the right axis of Figs. \ref{Fig3_MT_MH}(a) and (b) for the TbSi and TbSi$_{0.6}$Ge$_{0.4}$ samples, respectively. We fit the 1/$\chi$ vs. T curves at various magnetic fields [see Fig. 2 of \cite{SI} and the solid black lines in Figs. \ref{Fig3_MT_MH}(a, b) at 100~Oe] for both samples, using the Curie-Weiss law: $\chi$=C/(T-$\theta_{\rm p}$), where C and $\theta_{\rm p}$ are the Curie-Weiss constant and the paramagnetic Curie temperature, respectively, which gives an average value of $\theta_{\rm p}$= -7~K and -4~K, and effective magnetic moment ($\mu_{\rm eff.}$=$\sqrt{8\rm C}$)= 10.11 $\mu_B$/f.u. and 10.12 $\mu_B$/f.u. for TbSi and TbSi$_{0.6}$Ge$_{0.4}$, respectively. The almost field-independent values of the $\mu_{\rm eff.}$ [see Figs. 2(g, h) of \cite{SI}] are slightly higher than the theoretical value for the free Tb$^{3+}$ ion with $^7$F$_6$ ground term (9.72 $\mu_B$) for both samples, indicating the small but finite contribution of the itinerant moments from the conduction electrons in these samples. Furthermore, the negative values of $\theta_{\rm p}$'s confirm the AFM interactions in both samples. However, the values of $\theta_{\rm p}$ are significantly smaller than the T$_{\rm N1}$ for both samples, which is also observed in Refs. \cite{Papamantellos_JMMM_86, Papamantellos_JMMM_88}. This can be attributed to the deviation in the free Tb$^{3+}$ states due to the crystal field (CF) and hence non-negligible temperature-independent paramagnetic (TIP) susceptibility term either from transition within these CF split states or from the core electrons \cite{Mabbs_book_73}. Furthermore, we observe a small but consistent enhancement in $\theta_{\rm p}$ with an increase in the magnetic field for both samples [see Figs. 2(e, f) of \cite{SI}], suggesting the possibility of short-range magnetic correlations in these samples in the paramagnetic state.

The field-dependent magnetization (M-H) curves have been recorded at 2~K (in LT AFM phase) and 50~K (in HT AFM phase) for TbSi and TbSi$_{0.6}$Ge$_{0.4}$, as shown in Figs. \ref{Fig3_MT_MH}(c) and (d), respectively. Interestingly, the M-H curves at 2~K show linear behavior for both samples, whereas those at 50~K display a notable convex nature at higher magnetic fields, suggesting the emergence of a metamagnetic-like transition in both samples in the HT AFM region. In order to clearly present this, the dM/dH versus H curves have been plotted in the insets of Figs. \ref{Fig3_MT_MH}(c) and (d) for one full M-H cycle (-70~kOe to +70~kOe). A significant increase in the slope of the M-H curves up to around 60~kOe is likely due to the ongoing spin-reorientation process \cite{Zhuang_PRB_23, Kumar_PRB1_20}. The dM/dH decreases with further increase in H for H$>$60~kOe for both samples. The magnetization at 70~kOe and 50~K is $\sim$2.5~$\mu_B$/f.u. for both samples, which is around 28\% of the theoretical saturation value (M$_{\rm S} =$ g$_J$J = 9 $\mu_B$) for Tb$^{3+}$ ions, indicating that the magnetic state is still predominantly AFM. A metamagnetic transition with a critical field of 65~kOe has also been observed in DySi with 80\% FM alignment of the Dy$^{3+}$ moments at 90~kOe and 5~K \cite{Nirmala_JMMM_08}. The linear behavior of the M-H curves below T$_{\rm N2}$ on the other hand suggests stronger AFM interactions (in response to the external magnetic field) in the LT phase of both samples. Here, it is worth mentioning that both samples have planar AFM structure in the LT phase region, which is commensurate with the crystal lattice; whereas in the HT region, TbSi has sinusoidally modulated incommensurate magnetic structure [q=(q$_x$, q$y$, q$z$], and TbSi$_{0.6}$Ge$_{0.4}$ has 50\% commensurated [q=(0, 0, 1/2)] and 50\% incommensurated [q=(q$_x$, 0, q$_z$)] magnetic phases \cite{Papamantellos_JMMM_88, Papamantellos_JMMM_93}. This suggests that the modulation of the LT commensurated magnetic structures in the HT phase region decreases the strength of the AFM coupling in both samples. Moreover, both samples show negligible hysteresis in the M-H loop at 2~K as well as 50~K, which rules out the presence of any ground-state anisotropic FM correlations and/or domains.

\begin{figure}
\includegraphics[width=3.5in]{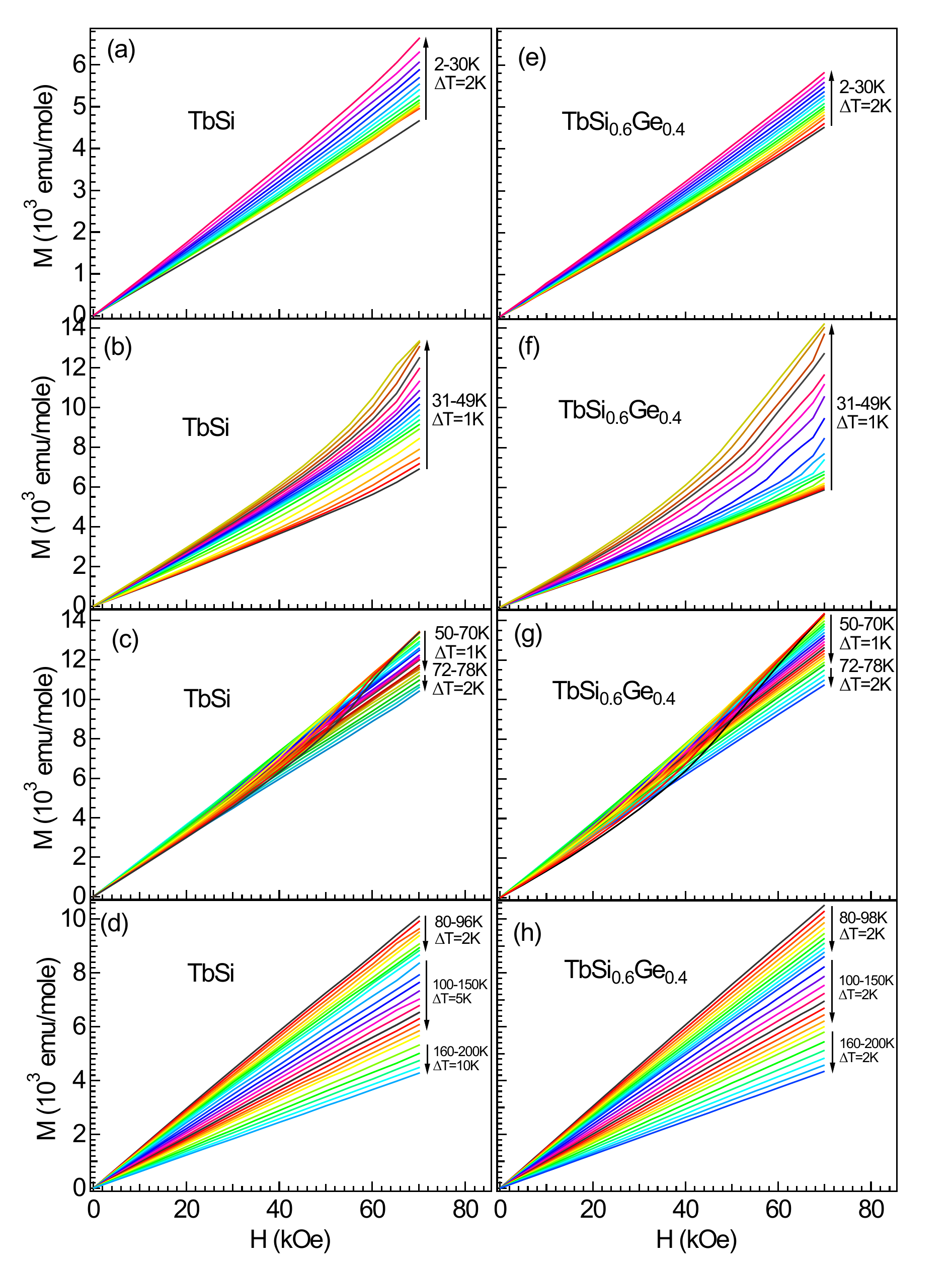}
\caption {The magnetization isotherms recorded from 0 to 70~kOe in the indicated temperature regimes for (a--d) TbSi and (e--h) TbSi$_{0.6}$Ge$_{0.4}$. } 
\label{Fig4_Virgin}
\end{figure}

Further, the magnetization isotherms have been recorded from 0 to 70~kOe across the transition temperatures, T$_{\rm N1}$ and T$_{\rm N2}$, and are shown in Figs. \ref{Fig4_Virgin}(a--d) for TbSi and \ref{Fig4_Virgin}(e--h) for TbSi$_{0.6}$Ge$_{0.4}$. It can be clearly observed that the virgin curves are sufficiently straight below 30~K for both samples [see Figs. \ref{Fig4_Virgin}(a, e)], indicating rigid coupling in the LT AFM phase in both samples. The curves begin to display a convex shape above 30~K, which becomes more pronounced as the temperature approaches 50~K, as shown in Fig. \ref{Fig4_Virgin}(b) for TbSi. However, the convex character of the virgin isotherm at 31~K in the case of TbSi$_{0.6}$Ge$_{0.4}$ is weaker compared to that of TbSi [see Figs. \ref{Fig4_Virgin}(b, f)], which confirms the better stability of the LT planar AFM in the former. Moreover, close to T$_{\rm N1}$, i.e., above 50~K, both samples show the ``S-like" shape of the virgin curves [see Figs. \ref{Fig4_Virgin}(c, g)], which is a typical signature of the metamagnetic transition \cite{Kumar_PRB_08}. This affirms the weaker AFM coupling in these samples in the HT phase region due to finite incommensuration in the magnetic structure for T$_{\rm N2} <T < $ T$_{\rm N1}$, discussed above \cite{Papamantellos_JMMM_88, Papamantellos_JMMM_93}. Above 50~K, the magnetization increases with the temperature for low magnetic fields but decreases monotonically at 70~kOe for both samples, as shown in Figs. \ref{Fig4_Virgin}(c, g), which indicates the shift in the T$_{\rm N1}$ to the lower temperature with an increase in the magnetic field. Figures \ref{Fig4_Virgin}(d, h) show the virgin curves for both samples deep in the paramagnetic region (T$\geqslant$80~K), where the curves exhibit a slight concave nature even up to around 100~K. Thus, the presence of short-range magnetic correlations is possible in these compounds. Recently, short-range magnetic interactions have been also detected in other AFM materials using Raman spectroscopy up to temperatures as high as $\sim$3T$_{\rm N}$ \cite{Singh_PRR_20}.

\begin{figure*}
\includegraphics[width=7in]{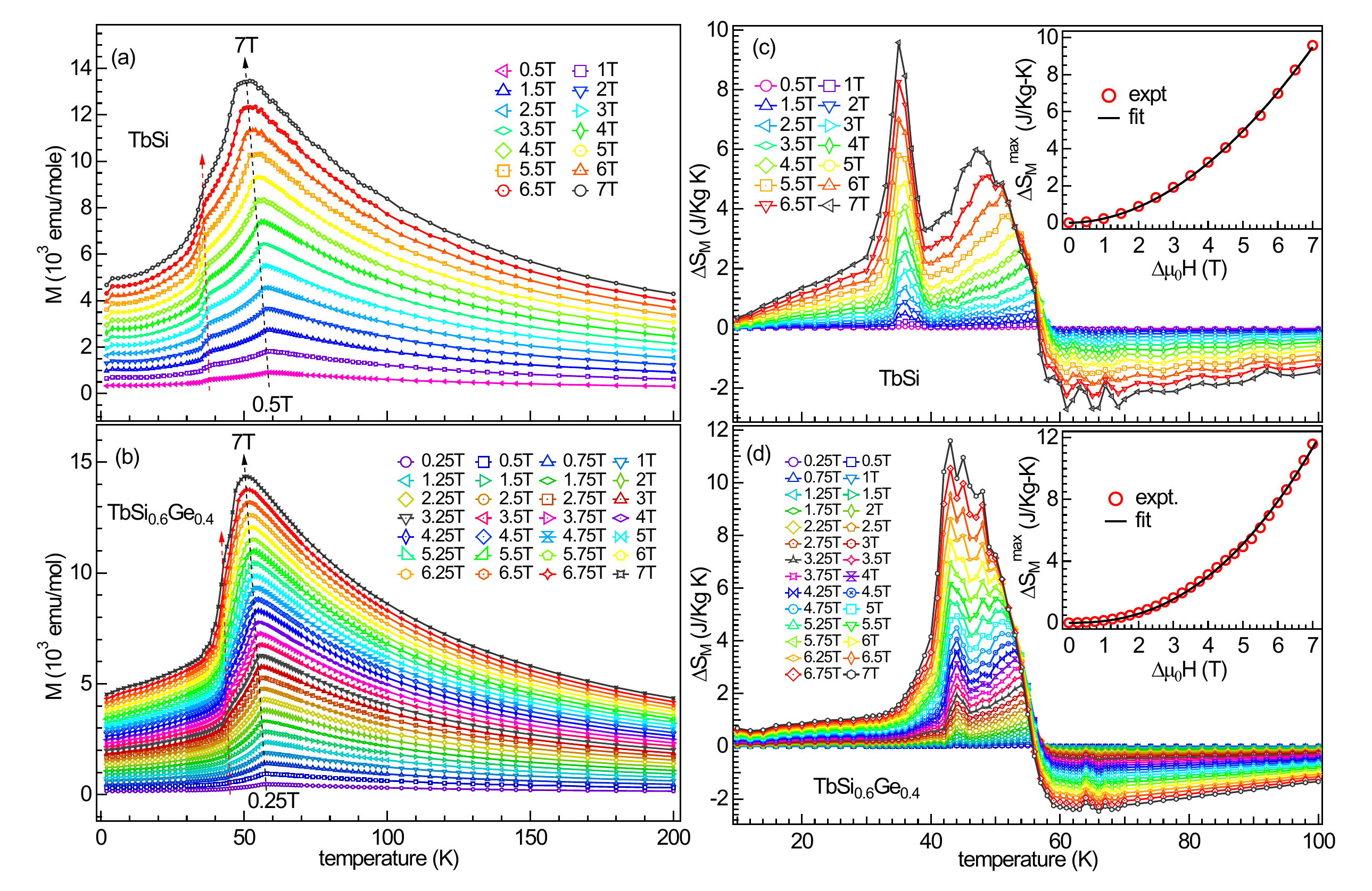}
\caption {Temperature dependent magnetization at different magnetic fields extracted from the isothermal curves for (a) TbSi and (b) TbSi$_{0.6}$Ge$_{0.4}$. Black and red dashed arrows represent the shift in T$_{\rm N1}$ and T$_{\rm N2}$, respectively,  with increase in magnetic field.  (c, d) The temperature dependent magnetic entropy change ($\Delta$S$_{\rm M}$) for different $\Delta \mu_0$H in the vicinity of  T$_{\rm N1}$ and T$_{\rm N2}$ for both the samples. Insets show the field dependent $\Delta S_{\rm M}^{max}$ at T$_{\rm N2}$, where black solid curves represent the best fit using the power law.} 
\label{Fig5_MCE}
\end{figure*}

The temperature-dependent magnetization (M-T) curves derived from these magnetization isotherms are plotted in Figs. \ref{Fig5_MCE}(a, b) for TbSi and TbSi$_{0.6}$Ge$_{0.4}$, respectively. It can be clearly observed that both transition temperatures, T$_{\rm N1}$ and T$_{\rm N2}$, shift to lower values with an increase in the magnetic field, as indicated by black and red dashed arrows, respectively. However, this effect is much more prominent for T$_{\rm N1}$ compared to T$_{\rm N2}$ [see Figs. \ref{Fig5_MCE} (a, b) and Figs. 2 (a, b) of \cite{SI} for more clarity], supporting weaker AFM coupling in the HT phase. In fact, we observe a significant broadening and hence a plateau-like feature at T$_{\rm N1}$ in the case of TbSi for H$>$5~T [see Fig. \ref{Fig5_MCE}(a)]. This effect is more clearly evident in the high-field specific heat measurements, discussed below. Moreover, in the case of TbSi, the transition at T$_{\rm N2}$ remains sharp and distinctly visible even at 7~T, whereas it becomes broad and nearly merges with T$_{\rm N1}$ in the case of TbSi$_{0.6}$Ge$_{0.4}$ [see Figs. \ref{Fig5_MCE} (a, b)]. This is due to the evolution of additional field-induced magnetic interactions between T$_{\rm N1}$ and T$_{\rm N2}$ in the latter, which we discuss below in detail.\par

The change in magnetic entropy ($\Delta$S$_{\rm M}$) is a crucial parameter for investigating the magnetocaloric effect (MCE) and can also serve as a sensitive tool to probe the interplay of complex magnetic interactions and their evolution with temperature and magnetic field in magnetically ordered compounds \cite{Kumar_PRB1_20, Kumar_PRB_08, Phan_JMMM_07, Wang_AEM_24}. Therefore, $\Delta$S$_{\rm M}$ is calculated as a function of temperature for different $\Delta$H values using the following Maxwell’s thermodynamic relation \cite{Phan_JMMM_07}.

\begin{eqnarray} 
 \Delta S_M(\Delta H, T)=\int_0^H\bigg (\frac{\partial M(T,H)} {\partial T}\bigg)_H dH
\end{eqnarray}

\begin{table*}
\label{Table_fit}
\caption{The comparison of the transition temperatures (T$_{\rm C}$/T$_{\rm N}$), $\Delta$S$_{\rm M}^{max}$, and TEC(10) of various Tb-based compounds with TbSi and TbSi$_{0.6}$Ge$_{0.4}$ for the given values of change in the magnetic field ($\Delta \mu_0$H).}

\begin{tabular}{p{4.8cm}p{2cm}p{2cm}p{3cm}p{3cm}p{2cm}}
\hline
\hline
Sample & T$_{\rm C}$/T$_{\rm N}$ (K) & $\Delta \mu_0$H(T) & $\Delta$S$_{\rm M}^{max}$ (J/Kg-K) & TEC(10) (J/Kg-K) & Reference \\
\hline
TbSi &  35 & 7 & 9.6 & 5.4 & present study  \\
TbSi$_{0.6}$Ge$_{0.4}$ &  44 & 7& 11.6 & 9.3 & present study \\
TbCr$_2$Si$_2$C &  16.6 & 7& -16.4 & 15.79 & \cite{Ma_JalCom_22}  \\
TbMn$_2$Ge$_2$ & 95 & 7$||c$ & -24.02 & 21.39 & \cite{Huang_SR_22} \\
Tb$_2$Dy$_4$FeSb$_2$ &  182 & 5& -7.72 & 7.45 & \cite{Herrero_DT_23}\\
Tb$_4$CoIn & 50 & 7 & -7.2 & 7.1 & \cite{Remya_JRE_23}  \\
GdTbDyHo & 198 & 5 & -8.2 & 8.11 & \cite{Wang_AEM_24}\\
Tb$_{1.4}$Dy$_{0.6}$In & 155 & 5 & -5.97 & 5.9 & \cite{Remya_JalCom_24}\\
Tb$_2$Co$_{0.8}$Si$_{3.2}$ & 58 & 7 & -3.32 & 3.2 & \cite{Remya_JPCM_24}\\
TbCo$_{2}$ & 233 & 5 & -5.7 & 5.4 & \cite{Bykov_JalCom_24}  \\
Gd$_{0.2}$Tb$_{0.2}$Dy$_{0.2}$Ho$_{0.2}$Er$_{0.2}$Ni$_2$ & 32 & 5 & -14.3 & 13.8 & \cite{Jesla_JMMM_24}\\
\hline
\hline
\end{tabular}
\end{table*}

The $\Delta$S$_{\rm M}$(T) curves for the TbSi and TbSi$_{0.6}$Ge$_{0.4}$ samples are depicted in Figs. \ref{Fig5_MCE}(c) and (d), respectively. A clear transition from the inverse magnetocaloric effect (IMCE) (positive $\Delta$S$_{\rm M}$) to the conventional magnetocaloric effect (CMCE) (negative $\Delta$S$_{\rm M}$) is observed at $T_{\rm N1} \approx$ 58 K (for $\Delta \mu_0$H = 0.5~T) for both samples, as expected for the AFM to PM transition \cite{Biswas_PRB_13}. This crossover temperature decreases with the increase in applied magnetic field as a consequence of a field-induced shift in $T_{\rm N1}$, as discussed above. Moreover, we observe a stronger peak in the $\Delta S_{\rm M}$ curves around 36 K (for $\Delta \mu_0$H = 0.5~T) corresponding to the LT AFM transition ($T_{\rm N2}$) in the case of the TbSi sample [see Fig. \ref{Fig5_MCE}(c)]. This large IMCE, with the value of $\Delta S_{\rm M}$ = 9.6 J/kg-K at $\Delta\mu_0$H = 7~T, indicates the possible use of TbSi as a heat sink to improve the efficiency of magnetic refrigerators \cite{proceeding}. Interestingly, a significant broadening in the $\Delta S_{\rm M}$ peak of TbSi$_{0.6}$Ge$_{0.4}$ has been observed at higher magnetic fields, where the two peaks corresponding to LT and HT AFM phases almost merge into each other at $\Delta\mu_0$H = 7~T [see Fig. \ref{Fig5_MCE}(d)]. This is due to the evolution of additional maxima in the $\Delta S_{\rm M}$ curves of TbSi$_{0.6}$Ge$_{0.4}$ between $T_{\rm N1}$ and $T_{\rm N2}$ at higher fields. The presence of maxima in the $\Delta S_{\rm M}$ curves, i.e., positive entropy change (IMCE), at higher fields indicates the evolution of field-induced additional AFM coupling(s) in the HT phase region of this sample. This is more clearly evident from the behavior of specific heat data at higher magnetic fields, discussed below. The maximum value of $\Delta S_{\rm M}$ ($\Delta S_{\rm M}^{max}$) at $\Delta\mu_0$H = 7~T increases to 11.6 J/kg-K in TbSi$_{0.6}$Ge$_{0.4}$, which along with the significant broadening in the $\Delta S_{\rm M}$ peak compared to pure TbSi [see Figs. \ref{Fig5_MCE}(c, d)], makes the TbSi$_{0.6}$Ge$_{0.4}$ sample more suitable for magnetic refrigerator applications. We note that the maximum MCE of our compounds is smaller compared to other well-known materials in this temperature range, such as Eu$_2$In \cite{Guillou_NC_18}, but the plateau-like temperature profile of the TbSi$_{0.6}$Ge$_{0.4}$ compound can make it enticing for engineers looking for materials with inverse MCE for hydrogen liquefaction. For instance, a rough approximation of a single peak corresponding to IMCE in the case of TbSi$_{0.6}$Ge$_{0.4}$ gives a moderate relative cooling power (RCP) of 125 J/kg for $\Delta \mu_0$H = 7~T, where RCP is defined as RCP = $\Delta S_M^{\rm max}\delta_{\rm FWHM}$, where $\delta_{\rm FWHM}$ represents the full width at half maximum of the peak. Further, we calculated the temperature-averaged entropy change (TEC), defined by Griffith et al. \cite{Griffith_JAP_18}, as 

\small
\begin{equation} 
{\rm TEC}(\Delta T_{\rm lift})=\frac{1}{\Delta T_{\rm lift} }\stackunder{max}{\scriptsize Tmid} \left\{\int\limits_{T_{\rm mid-\frac{\Delta T_{\rm lift}}{2}}}^{T_{\rm mid+\frac{\Delta T_{\rm lift}} {2}}}\Delta S_M(T)_{\Delta H, T}dT\right\}
\end{equation}
\normalsize

where, $\Delta T_{\rm lift}$ represents the temperature range within which the material exhibits its optimal response to $\Delta$H, while T$_{\rm mid}$ denotes the temperature at which ${\rm TEC}(\Delta T_{\rm lift})$ reaches its maximum value. This value is determined by sweeping over different temperature ranges for a given $\Delta T_{\rm lift}$. We obtained a reasonably high value of TEC(10) = 9.3~J/kg-K for TbSi$_{0.6}$Ge$_{0.4}$ at T$_{\rm mid}$ = 46~K, compared to TEC(10) = 5.4~J/kg-K for TbSi at T$_{\rm mid}$ = 37~K, both at $\Delta\mu_0$H = 7~T, owing to the relatively broad transition in the former. These $\Delta S_{\rm M}^{max}$ and TEC(10) values are moderate compared to other Tb-based compounds listed in Table I. However, such values, along with positive $\Delta S_{\rm M}$ (IMCE) and negligible thermal hysteresis, make these samples particularly significant, considering the low magnetic moment and sharp nature of AFM transitions, resulting in small values of $\Delta S_{\rm M}$ and TEC(10), respectively, in materials exhibiting IMCE. Furthermore, the values of $\Delta S_{\rm M}^{max}$ at T$_{\rm N2}$ monotonically increase with the increase in $\Delta \mu_0$H up to 7~T for both samples. The field dependence of $\Delta S_{\rm M}^{max}$ is depicted in the insets of Fig. \ref{Fig5_MCE}(c, d) for TbSi and TbSi$_{0.6}$Ge$_{0.4}$, respectively. The best fit of the $\Delta S_M^{max}$($\Delta$H) curves using the power law $\Delta S_{\rm M}^{max}\propto$H$^\phi$ is illustrated by the solid black curves, yielding $\phi =$1.93(2) and 2.36(4) for TbSi and TbSi$_{0.6}$Ge$_{0.4}$, respectively. The nearly quadratic H dependence of $\Delta S_{\rm M}^{max}$ and the rigidity of the LT AFM phase in response to the external field suggest that the giant magnetocaloric effect can be achieved in these samples at the higher magnetic fields \cite{Gottschall_PRB_19}. For instance, extrapolation of the $\Delta S_{\rm M}^{max}$($\Delta$H) curves gives $\Delta S_{\rm M}^{max}$ = 36.1 and 58.6 J/kg-K at $\Delta\mu_0$H = 14~T for TbSi and TbSi$_{0.6}$Ge$_{0.4}$, respectively.

\subsection{Specific heat}

\begin{figure}
\centering
\includegraphics[width=3.5in]{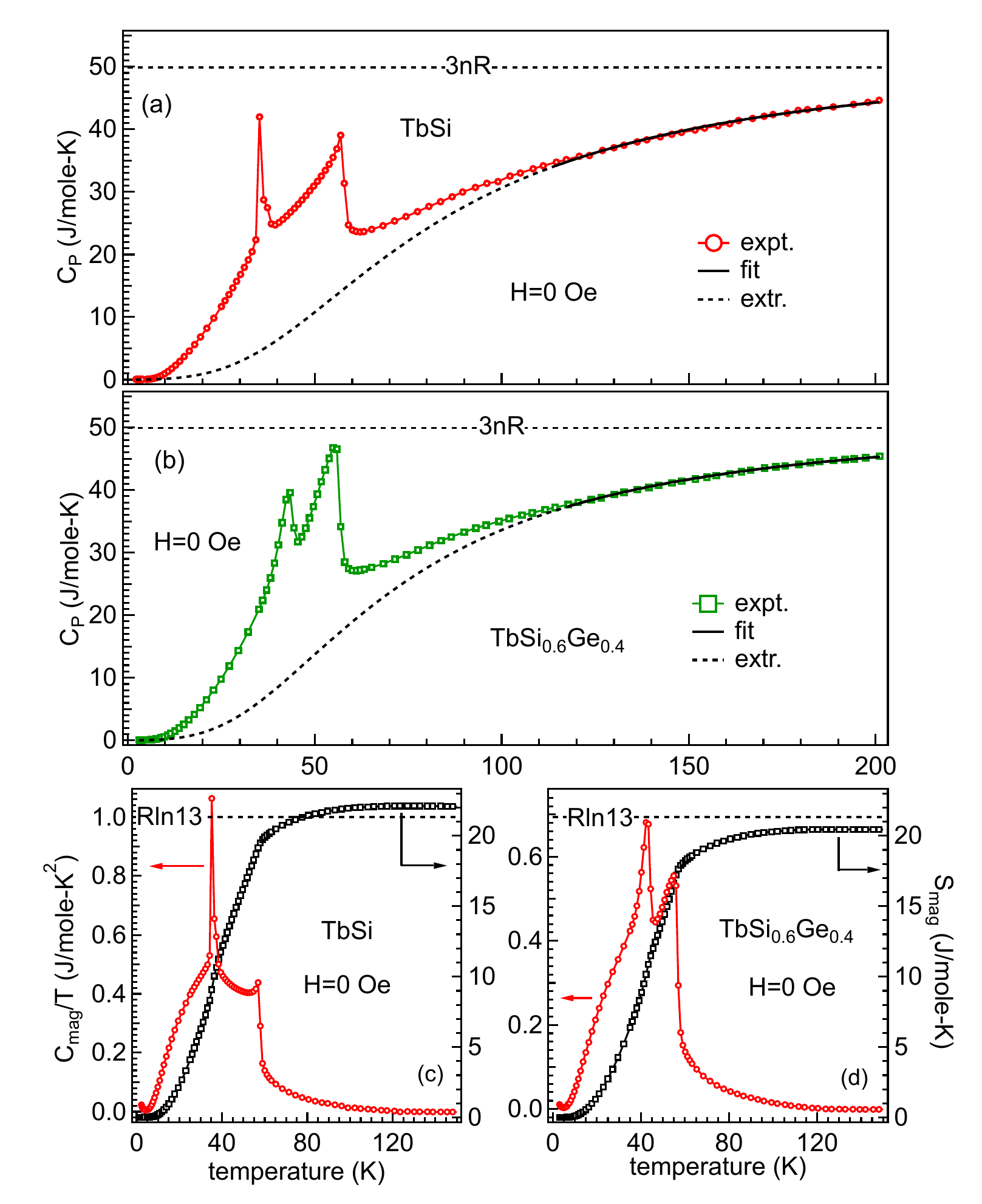}
\caption {The zero-field temperature-dependent specific heat curves measured using the standard relaxation technique for (a) TbSi and (b) TbSi$_{0.6}$Ge$_{0.4}$ samples. The solid black curves represent the best fit obtained using the sum of electronic and Debye lattice specific heat from 115-200~K, with dashed curves indicating their extrapolation down to $\sim$2~K. The horizontal dashed lines denote the classical Dulong-Petit limit. (c, d) The temperature-dependent magnetic specific heat (C$_{\rm mag}$)/T (on the left axis) and magnetic entropy (S$_{\rm mag}$) (on the right axis) for the TbSi and TbSi$_{0.6}$Ge$_{0.4}$ samples, respectively. Dashed lines represent the theoretical value of S$_{\rm mag}$, Rln13, for the free Tb$^{3+}$ ion. } 
\label{Fig6_HC_0T}
\end{figure}

The substantial IMCE and TEC(10) at T$_{\rm N2}$ and the noteworthy enhancement in the latter with the Ge substitution, make it imperative to investigate the nature of this LT AFM transition in these samples.  Therefore, we record the zero-field specific heat data from 2 to 200~K, as depicted in Figs. \ref{Fig6_HC_0T}(a, b) for TbSi and TbSi$_{0.6}$Ge$_{0.4}$, respectively. Two well-defined peaks have been observed at T$_{\rm N2}$ $\approx$ 35~K and 43~K, and T$_{\rm N1}$ $\approx$ 57~K and 56~K for the TbSi and TbSi$_{0.6}$Ge$_{0.4}$ samples, respectively, which is consistent with the  LT and HT AFM orderings observed in the magnetization measurements. We observe an asymmetric $\lambda$-like peak at T$_{\rm N1}$ for both both TbSi and TbSi$_{0.6}$Ge$_{0.4}$, suggesting the second-order nature of this HT AFM to PM transition in  these samples \cite{Kumar_PRB2_20}. On the other hand, the peak at T$_{\rm N2}$ is delta-like, sharp and symmetric in nature, indicating the first-order nature of the LT AFM to HT AFM phase transition \cite{Pecharsky_JAP_99}. The specific heat at 200~K at is around 45~J/mole-K for both  samples, which is slightly lower than the classical Dulong-Petit limit, i.e., 3$n$R=49.9~J/mole-K, where $n$ and R are the number of atoms per formula unit (two in this case) and  gas constant, respectively, as shown by horizontal dashed lines in Figs. \ref{Fig6_HC_0T}(a, b).  We estimate the lattice and electronic contribution in the specific heat curves by fitting the C$_P$(0, T) curves from 115-200~K using the Debye model with a single Debye temperature, and an electronic term, $\gamma$T, as given by \cite{Kittel_book_05}

\begin{equation}
C_V (T) = \gamma T +  9nR\left(\frac{T}{\theta_D}\right)^3 \int_0^{\theta_D/T} \frac{x^4e^x}{(e^x-1)}dx,
\label{Debye_HC}
\end{equation} 

\begin{figure}
\centering
\includegraphics[width=3.5in]{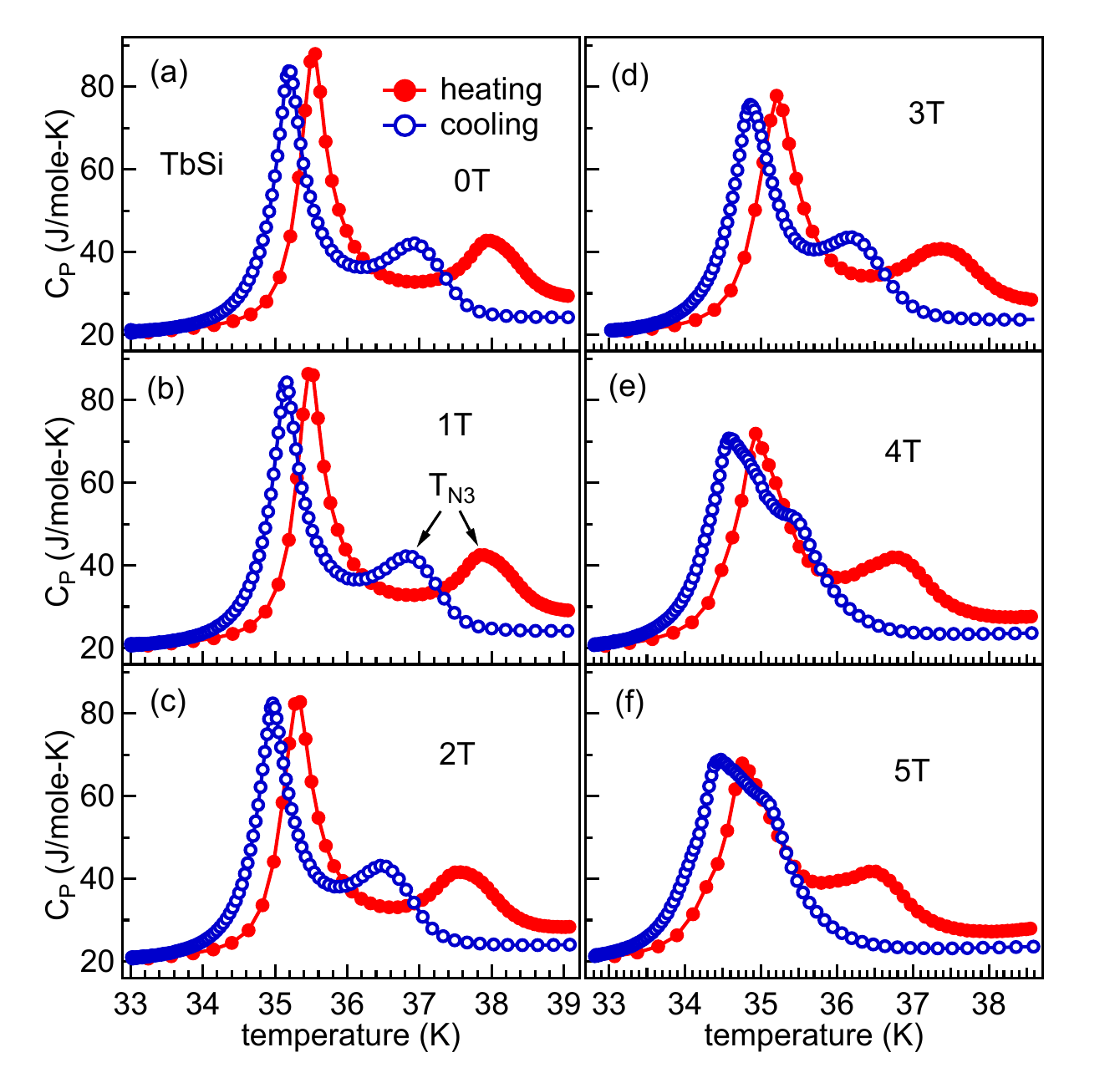}
\caption {(a--f) The temperature dependent specific heat curves of TbSi measured using heat pulse method in the vicinity of T$_{\rm N2}$ in both heating and cooling modes at different magnetic fields from 0--5~T.  Arrows in (b) mark the additional transition (IT AFM) observed in the high resolution C$_P$ data. } 
\label{Fig7_HC_TbSi1}
\end{figure}

\begin{figure}
\centering
\includegraphics[width=3.5in]{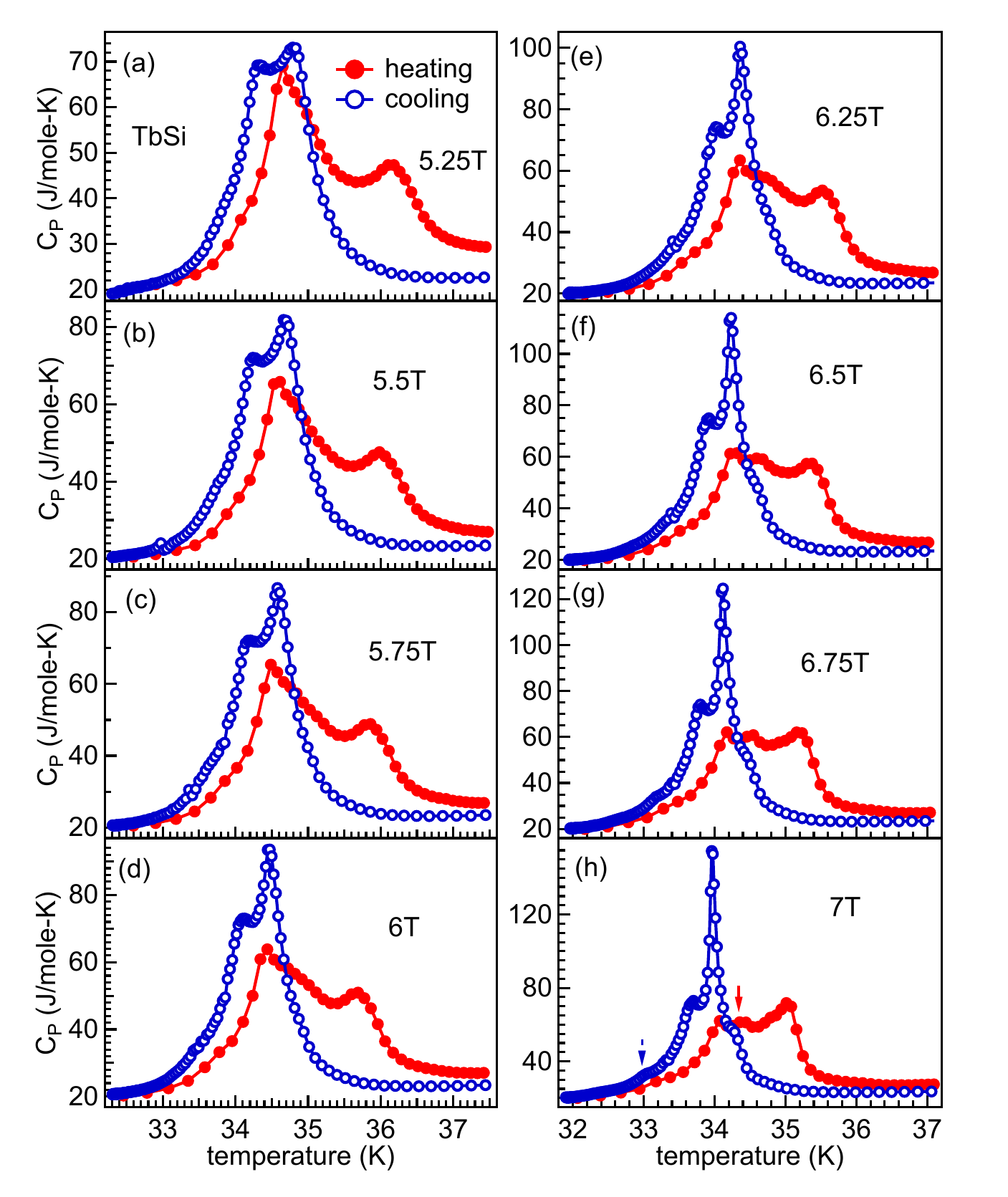}
\caption {(a--h) The temperature dependent specific heat curves of TbSi in the vicinity of T$_{\rm N2}$ in both heating and cooling modes for $\mu_0$H= 5.25--7~T. Arrows in (h) represent the evolution of addition features with increase in the magnetic field.  } 
\label{Fig8_HC_TbSi2}
\end{figure}

where $\theta_D$, and $\gamma$ are the Debye temperature and Sommerfeld coefficient, respectively. The solid black curves in Figs. \ref{Fig6_HC_0T}(a) and (b) depict the best fit using the Debye function, which gives $\theta_D$ = 333$\pm$3~K and 296$\pm$2~K,  and $\gamma$ = 3.4(6) and 2.1(3) mJ/mole-K$^2$ for the TbSi and TbSi$_{0.6}$Ge$_{0.4}$ samples, respectively. We extrapolate these curves down to $\sim$2~K, as indicated by the black dashed curves in Figs. \ref{Fig6_HC_0T}(a, b), to extract the lattice and electronic contribution across the entire temperature range under study. The magnetic entropy (S$_{mag}$) is then calculated for both samples using the formula S$_{mag}$(T) = $\int \frac{C_{mag}}{T}$ dT, where C$_{mag}$ represents the magnetic specific heat obtained by subtracting the sum of lattice and electronic specific heat (dashed lines) from the C$_{\rm P}$(0, T) curves. The temperature-dependent C$_{mag}$/T and S$_{mag}$ are shown on the left and right axes of Figs. \ref{Fig6_HC_0T}(c) and (d) for the TbSi and TbSi$_{0.6}$Ge$_{0.4}$ samples, respectively. The S$_{mag}$ exhibits a saturation value of around 20~J/mole-K for both samples, close to the theoretical value of Rln(2J+1) = Rln13 = 21.33~J/mole-K for a free Tb$^{3+}$ ion (J=6), as illustrated by the dashed lines in Figs. \ref{Fig6_HC_0T}(c) and (d). However, a precise estimation of the lattice specific heat using non-magnetic isostructural reference samples is important for a more reliable estimation of S$_{mag}$.

Further, we record the temperature-dependent specific heat curves at different magnetic fields [C$_{\rm P}$(H, T)] for both TbSi and TbSi$_{0.6}$Ge$_{0.4}$ samples. We observe that the low-temperature peak (at T$_{\rm N2}$) remains sharp and almost symmetric up to 7~T [see Figs. 3(a,b) of \cite{SI}], which confirms the first-order nature of the LT to HT AFM transition in both samples. Moreover, the TbSi displays an additional sharp peak in the vicinity of the LT AFM, as shown by the arrow in Fig. 3(a) of \cite{SI}. Here, it is important to mention that the specific heat measurements performed using the conventional thermal-relaxation technique assume that C$_{\rm P}$ remains constant, and have the same values during heating and cooling cycles for a given heat pulse. However, for a very sharp (delta-like) FOPT, as in the present case at T$_{\rm N2}$, both of these assumptions become unreliable, leading to considerable inaccuracies in the measured values of C$_{\rm P}$ in the vicinity of FOPT \cite{Guillou_NC_18, Hardy_JPCM_09, Lashley_Cryo_43}. Therefore, a point-by-point time-dependent analysis of both heating and cooling curves, after giving a long heat pulse which can essentially capture the whole phase transition, is more appropriate in the present case \cite{Lashley_Cryo_43}. In Figs. \ref{Fig7_HC_TbSi1}(a--f) and \ref{Fig8_HC_TbSi2}(a--h), we show the specific heat curves of TbSi measured using this method in the vicinity of LT AFM in both heating and cooling cycles at different applied magnetic fields. To record the data at each field, the sample was first heated up to 100~K and then cooled down to the desired temperature in the absence of magnetic field, in order to remove the possibility of any thermal and/or magnetic remanence. The temperature and magnetic field were then stabilized for 20 minutes before applying a heat pulse for 500~secs. A clear shift of $\sim$0.3~K in T$_{\rm N2}$ measured in heating and cooling modes can be observed at all the magnetic fields, which confirm the first-order nature of this LT AFM to HT AFM phase transition in TbSi. This small thermal hysteresis was overshadowed due to the larger temperature step (1~K) used in the magnetization measurements performed in heating and cooling modes, discussed above. This demonstrates the potential of the C$_{\rm P}$ measurements performed using the heat pulse method in resolving the fairly sharp features in the vicinity of FOPT. Furthermore, the peak value of C$_{\rm P}$ significantly increases compared to that observed using the standard relaxation technique [see Figs. \ref{Fig6_HC_0T}(a) and \ref{Fig7_HC_TbSi1}(a)], as the instantaneous value of C$_{\rm P}$ in the former is free from both of the assumptions described above \cite{Guillou_NC_18, Hardy_JPCM_09}. More importantly, now we clearly see an additional somewhat broad but almost symmetric feature above T$_{\rm N2}$ in both heating and cooling modes, represented as T$_{\rm N3}$ in Fig. \ref{Fig7_HC_TbSi1}(b). In the absence of magnetic field, this feature has almost half intensity but significantly higher thermal hysteresis ($\sim$2~K) as compared to the LT AFM transition, indicating the first-order nature of this transition as well. Here, it is important to reemphasize that ND measurements performed on the TbSi showed the planar (collinear) nature of the LT AFM phase below 36~K [q=(0, 1/2, 0)], where the magnetic moments lie in the $xy$-plane, making an angle of 66$\pm$2$^0$ with the $a$-axis \cite{Papamantellos_JMMM_93}. This LT planar AFM was found to transformed into a three dimensionally modulated incommensurate HT AFM phase with q=(q$_x$, q$_y$, q$_z$) \cite{Papamantellos_JMMM_93}. However, the most striking modulation in the magnetic wave vector was observed between 36-39~K, and above 42~K magnetic structure is almost commensurate with the crystal lattice with q$\approx$(0, 1/2, 1/8) \cite{Papamantellos_JMMM_93}. The presence of additional peak in the intermediate temperature (IT) range ($\sim$36-39~K on heating) in the C$_{\rm P}$(0, T) curves clearly indicates the existence of a distinct magnetic phase between LT and HT AFM phases of TbSi. This peak shifts to the lower temperature with increase in the magnetic field [see Figs. 4(a, c) and 5(a, c) of \cite{SI}], which indicates the AFM nature of this phase. Therefore, in TbSi, the LT planar AFM transformed into an IT AFM via a FOPT, which then transformed into HT AFM through another FOPT and finally into PM state with second order phase transformation (SOPT). In fact, an abrupt variation in the position, line width, and the integrated intensity of (1, 1/2, 0) reflection in the ND data reported between 36-39~K in Ref. \cite{Papamantellos_JMMM_93} along with a thermal hysteresis of 2~K in TbSi is most likely to be associated with this IT AFM phase. In order to further confirm the nature of the IT to HT AFM phase transition, we apply a short heat pulse only up to $\sim$37~K (below T$_{\rm N3}$) and interestingly, no peak corresponding to IT AFM was observed in the cooling curve (see Fig. 6 of \cite{SI}), which confirms the first-order nature of this transition. Note that a slightly higher background of the heating curves at the higher temperatures in Figs. \ref{Fig7_HC_TbSi1} and \ref{Fig8_HC_TbSi2} originates from the partial breakdown of the ideal adiabatic condition at the end of the long heating pulse. The field induced shift in T$_{\rm N3}$ is higher than T$_{\rm N2}$, which causes an increase in the strength of IT AFM peak with the magnetic field, especially for $\mu_0$H$>$5~T, due to their overlap with the LT AFM peaks. This effect is more prominent in the cooling curves owing to their lesser separation ($\sim$1.7~K at 0~T) between LT and IT AFM peaks as compared to the heating curves ($\sim$2.5~K at 0~T) [see Figs. \ref{Fig8_HC_TbSi2} (a--h]. However, both transitions remain sharp even up to 7~T, where both heating and cooling curves confine within a narrow temperature range of $\sim$3~K, contrary to TbSi$_{0.6}$Ge$_{0.4}$, which shows significant broadening at the higher fields (discussed below). Moreover, the LT AFM peaks measured in the heating mode show an asymmetry towards higher temperatures for $\mu_0$H$>$5~T and a clear evolution of an additional feature for $\mu_0$H$>$6~T [see Figs. \ref{Fig8_HC_TbSi2}(a--h)], as shown by the solid red arrow in Fig. \ref{Fig8_HC_TbSi2}(h). However, due to their sharp and intense nature, this feature is visible in the cooling curves only for $\mu_0$H$>$6~T [see Figs. \ref{Fig8_HC_TbSi2} (e--h)]. Further, a weak feature is also deconvoluted below T$_{\rm N2}$ (between $\sim$33--34~K) in both heating and cooling curves for $\mu_0$H$>$5~T, as represented by the dashed blue arrow in Fig. \ref{Fig8_HC_TbSi2}(h) [see Fig. 5(b, d) of \cite{SI} for clarity]. Both of these field-induced features also shift to lower temperatures with an increase in the magnetic field, indicating their AFM nature. More importantly, both of these features evolve at the expense of the LT AFM peak [see Figs. \ref{Fig8_HC_TbSi2}(a--h)] and show a thermal hysteresis of $\sim$0.3~K. This indicates the field-induced modulation of the LT AFM phase into different AFM configurations having small formation energy differences as their origin \cite{Feng_PRB_13}. We call these low- and high-temperature field-induced features split from the LT AFM as (LT AFM)$_{\rm S1}$ and (LT AFM)$_{\rm S2}$, respectively. Furthermore, we calculate the magnetic entropy and adiabatic temperature change ($\Delta$T$_{\rm ad}$) using these C$_{\rm P}$(H, T) curves for TbSi [shown in Figs. 7(a--d) of \cite{SI}], which show good agreement (in $\Delta$S$_{\rm M}$) with those calculated from the magnetization measurements, discussed above.

\begin{figure}
\centering
\includegraphics[width=3.5in]{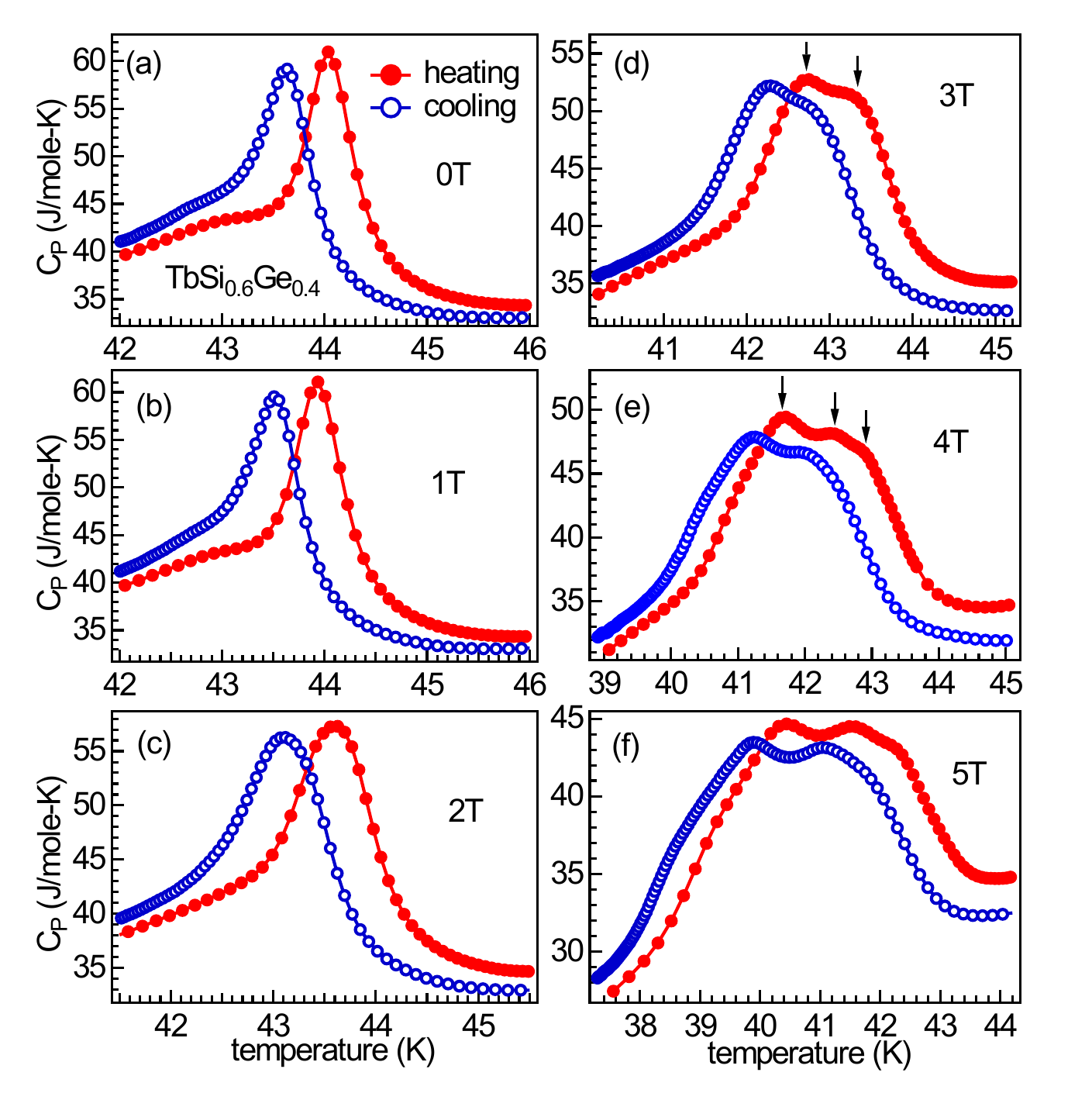}
\caption {(a--f) The temperature dependent specific heat curves of TbSi$_{0.6}$Ge$_{0.4}$ in the vicinity of T$_{\rm N2}$ in both heating and cooling modes at different magnetic fields from 0--5~T. Arrows in (d) and (e) show the splitting of the peaks into a doublet and triplet at 3~T and 4~T magnetic fields, respectively. } 
\label{Fig9_HC_TbSiGe1}
\end{figure}

\begin{figure}
\centering
\includegraphics[width=3.5in]{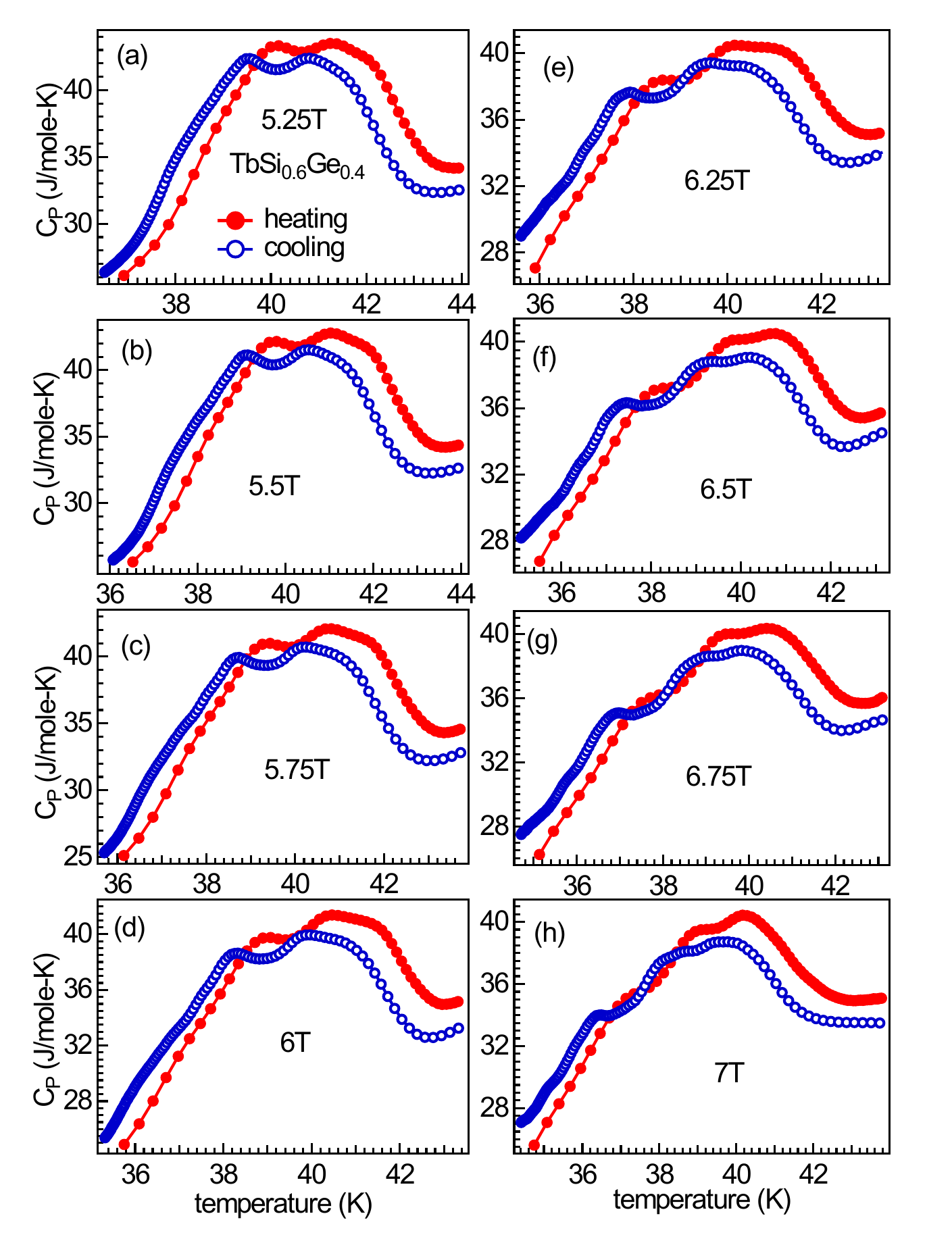}
\caption {(a--h) The temperature dependent specific heat curves of TbSi$_{0.6}$Ge$_{0.4}$ in the vicinity of T$_{\rm N2}$ in both heating and cooling modes for $\mu_0$H= 5.25--7~T.  } 
\label{Fig10_HC_TbSiGe2}
\end{figure}

\begin{figure*} 
\includegraphics[width=0.9\textwidth]{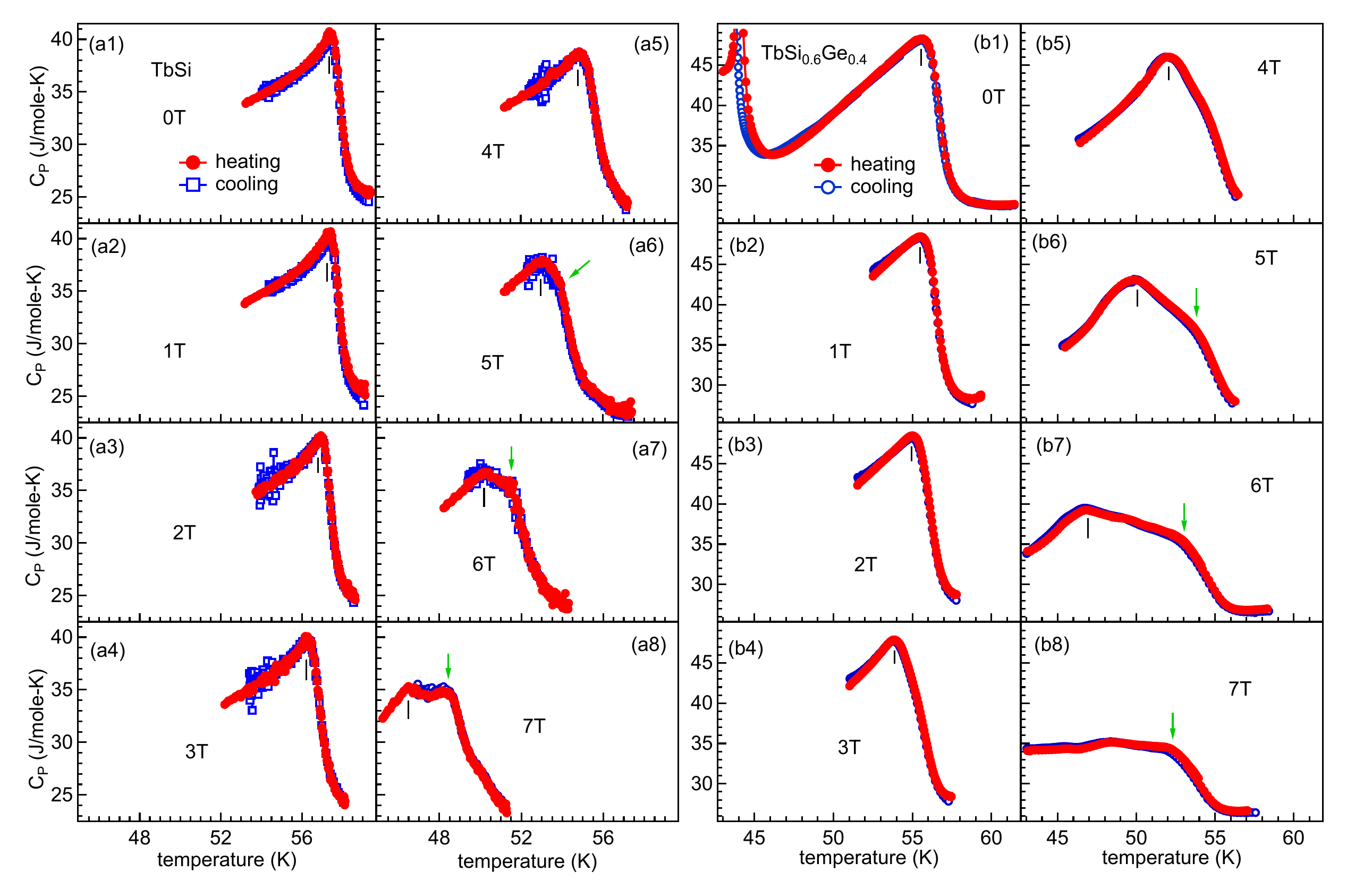}
\caption {The temperature dependent specific heat of (a1--a8) TbSi and (b1--b8) TbSi$_{0.6}$Ge$_{0.4}$ in the vicinity of T$_{\rm N1}$ at different magnetic fields. Black bars and green arrows represent the position of T$_{\rm N1}$ and the evolution of additional features for $\mu_0$H$\geqslant$5~T, respectively, in both the samples.}
\label{Fig11_HC_HT}
\end{figure*}

The TbSi$_{0.6}$Ge$_{0.4}$ sample, on the other hand, has an equal proportion of uniaxial AFM (collinear along the $c$-axis) with q=(0, 0, 1/2) and a two-dimensionally modulated (sinusoidal) AFM with q=(q$_x$, 0, q$_z$) in the HT AFM phase \cite{Papamantellos_JMMM_88}. Furthermore, the LT planar AFM phase (in the $ac$ plane) can also have the possibility of both collinear as well as canted alignment of the magnetic moments \cite{Papamantellos_JMMM_88}, and hence is expected to display an even more complex H-T phase diagram due to field-induced transformation between these different possible magnetic structures. The temperature-dependent specific heat curves of TbSi$_{0.6}$Ge$_{0.4}$ at the different magnetic fields from 0--5~T in the vicinity of LT AFM are shown in Figs. \ref{Fig9_HC_TbSiGe1}(a--f) in both heating and cooling modes. A hysteresis of around 0.4~K is observed in the C$_{\rm P}$(0, T) curves recorded in heating and cooling modes, indicating the first-order nature of the LT to HT AFM transition. However, we observe a shoulder-like feature towards the lower temperature side of the peak in both heating and cooling curves of this sample [see Fig. \ref{Fig9_HC_TbSiGe1}(a)]. The transformation of LT AFM into two different HT AFM phases [commensurate with q=(0, 0, 1/2) and incommensurate with q=(q$_x$, 0, q$z$) \cite{Papamantellos_JMMM_88}] may give rise to this shoulder-like feature in the C$_{\rm P}$(0, T) curves when one phase transformation is favored over the other. Furthermore, we observe a significant broadening in the LT AFM peak with an increase in the magnetic field even at 2~T [see Fig. \ref{Fig9_HC_TbSiGe1}(c)]. Interestingly, with a further increase in the magnetic field, this peak first splits into a doublet at $\mu_0$H=3~T and then into a triplet for $\mu_0$H$\geqslant$4~T, as shown by arrows in Fig. \ref{Fig9_HC_TbSiGe1}(d) and (e), respectively. All three peaks shift to lower temperatures with an increase in the magnetic field [see Figs. 8(a, c) and 9(a, c) of \cite{SI}], indicating their AFM nature. We define these low- and high-temperature field-induced features as (LT AFM)$_{\rm S1}$ and (LT AFM)$_{\rm S2}$, respectively. Importantly, the strength of (LT AFM)$_{\rm S1}$ and (LT AFM)$_{\rm S2}$ increases compared to the main LT AFM peak with an increase in the magnetic field. The field-induced increment in the latter is more prominent compared to the former. This effect in the C$_{\rm P}$(H, T) curves of TbSi$_{0.6}$Ge$_{0.4}$ is more clearly observed at higher fields for $\mu_0$H$>$5~T, as shown in Figs. \ref{Fig10_HC_TbSiGe2}(a--h). The strength of (LT AFM)$_{\rm S1}$ dominates over the main LT AFM peak for $\mu_0$H$\geqslant$5.25~T. Furthermore, the (LT AFM)$_{\rm S2}$ peak dominates over that of (LT AFM)$_{\rm S1}$ for $\mu_0$H$\geqslant$6.5~T [see Figs. \ref{Fig10_HC_TbSiGe2}(f--h)]. These observations indicate that with an increase in the magnetic field, the fraction of LT planar AFM transforming into uniaxial or 2D modulated AFM at zero field \cite{Papamantellos_JMMM_88} decreases, and that into two new field-induced HT AFM structures increases. Although with varying degrees, but all these LT to HT phase transitions coexist in TbSi$_{0.6}$Ge$_{0.4}$ for $\mu_0$H=4-7~T. The coexistence of these multiple phase transitions at the higher magnetic field in the case of TbSi$_{0.6}$Ge$_{0.4}$ causes the observed significant broadening in the $\Delta$S$_{\rm M}$ peaks (discussed above) compared to the pure TbSi sample. The thermal hysteresis has been observed for all these LT to HT AFM transitions, indicating their first-order nature.\par

\begin{figure*}
\centering
\includegraphics[width=0.9\textwidth]{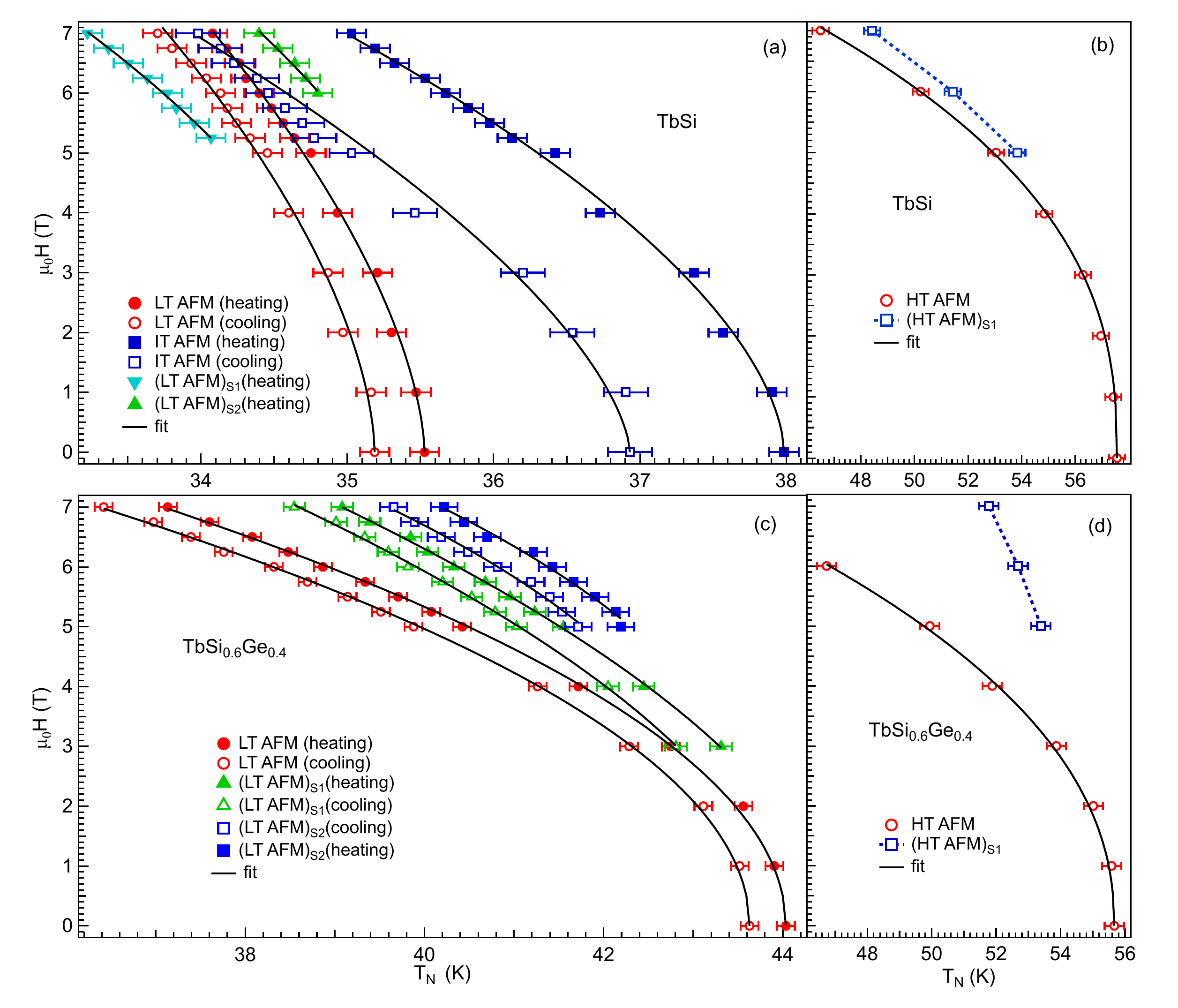}
\caption {The field dependent shift in the various AFM transition temperatures measured in heating and cooling modes for (a, b) TbSi and (c, d) TbSi$_{0.6}$Ge$_{0.4}$. The solid black curves represent the best fit using eq. \ref{decay}. } 
\label{Fig12_phase}
\end{figure*}

\begin{table*}
\label{Table_fit}
\caption{The best fit parameters extracted by fitting the field dependent AFM transition temperatures, extracted from the C$_P$(H, T) curves, using eq. \ref{decay} for the TbSi and TbSi$_{0.6}$Ge$_{0.4}$ samples.}

\begin{tabular}{p{3cm}p{5cm}p{3cm}p{3cm}p{3cm}}
\hline
\hline
& & T$_{\rm N0}$ (K) & $\gamma$ & $\mu_0$H$_{0}$ (T) \\
\hline
&  HT AFM & 57.6(1)& 0.39(3)& 13.5(8)   \\
& IT AFM (heating) & 38.0(1)& 0.58(6)& 31(4) \\
&  IT AFM (cooling) & 36.9(1)& 0.64(7)& 35(5) \\
TbSi &  LT AFM (heating) & 35.5(1)& 0.61(4)& 50(2) \\
&  LT AFM (cooling) & 35.2(1)& 0.59(5)& 47(3) \\
&  (LT AFM)$_{\rm S1}$ (heating) & 35.1(4)& 0.50(2)& 30(1) \\
&  (LT AFM)$_{\rm S2}$ (heating)& 35.9(4)& 0.50(3)& 34(2) \\
\hline
&  HT AFM & 55.7(1)& 0.45(2)& 13.8(4)   \\
&  LT AFM (heating) & 44.0(1)& 0.50(2)& 18(1) \\
&  LT AFM (cooling) & 43.6(1)& 0.50(2)& 17(1) \\
TbSi$_{0.6}$Ge$_{0.4}$ &  (LT AFM)$_{\rm S1}$ (heating) & 44.5(2)& 0.57(3)& 23(2) \\
&  (LT AFM)$_{\rm S1}$ (cooling)& 43.9(2)& 0.55(3)& 22(2) \\
&  (LT AFM)$_{\rm S2}$ (heating) & 43.9(2)& 0.40(2)& 19(2) \\
&  (LT AFM)$_{\rm S2}$ (cooling)& 43.4(3)& 0.40(3)& 18(2) \\
\hline
\hline
\end{tabular}
\end{table*}

Further, we also record the specific heat curves of both samples in the vicinity of the HT AFM to PM transition using the long heat pulse method, as shown in Figs. \ref{Fig11_HC_HT}(a1-a8) for TbSi and (b1-b8) for TbSi$_{0.6}$Ge$_{0.4}$, at different magnetic fields in both heating and cooling modes. No hysteresis has been observed in the C$_{\rm P}$ data recorded in heating and cooling modes, which is in agreement with the second-order nature of this transition in both samples. Interestingly, both samples show the evolution of an additional feature above T$_{\rm N1}$ for $\mu_0$H$\geqslant$5~T, as shown by solid green arrows in Figs. \ref{Fig11_HC_HT}(a6-a8) and \ref{Fig11_HC_HT}(b6-b8) for TbSi and TbSi$_{0.6}$Ge$_{0.4}$, respectively. The strength of this feature increases, accompanied by a shift to lower temperature with an increase in the magnetic field, causing a plateau-like behavior in the case of TbSi at 7~T, similar to that observed in the M-T measurements, discussed above. This indicates that the external magnetic field also stabilizes an AFM structure having a higher transition temperature compared to the zero-field HT AFM phase in both samples. This effect is more prominent in the case of TbSi$_{0.6}$Ge$_{0.4}$, where the main HT AFM peak almost merges into the LT AFM at 7~T. We use several long heat pulses at different temperatures to record the high-resolution C$_{\rm P}$ data in the wide temperature range in the case of TbSi$_{0.6}$Ge$_{0.4}$ for $\mu_0$H=0 and 4-7~T.\par
In order to quantitatively summarize these multiple magnetic interactions, we plot the field dependence of different transition temperatures [T$_{\rm N}$(H)] for TbSi and TbSi$_{0.6}$Ge$_{0.4}$ in Figs. \ref{Fig12_phase}(a, b) and (c, d), respectively.  The precise peak positions of different features have been determined by fitting the sum of Voigt functions. The error bars represent the estimated upper limit of the uncertainty, cumulatively from the peak fitting, temperature step size, and instrumental resolution. We fit the T$_{\rm N}$(H) data using the following relation \cite{Anand_PRB_16}

\begin{eqnarray} 
H= H_0\left(1-\frac{T_{\rm N}}{T_{\rm N0}}\right)^\gamma,
\label{decay}
\end{eqnarray}

where H$_0$ and T$_{\rm N0}$ represent the critical magnetic field required to shift T$_{\rm N}$ at 0~K and the AFM transition temperature at zero field, respectively. The solid black curves in Figs. \ref{Fig12_phase}(a--d) represent the best fit using the above equation, and the best fit parameters are listed in Table II. It can be clearly observed from Fig. \ref{Fig12_phase}(a) and (c) that the field-induced shift in the LT AFM is significantly higher in the case of TbSi$_{0.6}$Ge$_{0.4}$ compared to TbSi, which is also evident from the considerably lower values of H$_0$ for the former in both heating and cooling modes (see Table II). The difference between heating and cooling curves remains almost independent of the applied magnetic field for all the transitions in both samples. However, the separation of the field-induced features (LT AFM)$_{\rm S1}$ and (LT AFM)$_{\rm S2}$ from LT AFM peaks increases with the magnetic field for both samples. The higher overlapping of the (LT AFM)$_{\rm S1}$ and (LT AFM)$_{\rm S2}$ with LT AFM and IT AFM transitions, respectively, in the cooling curves of TbSi makes us unable to precisely determine their positions and hence they are not included in this H-T phase diagram. The field-dependent shift indicates the predominating AFM nature of all the transitions. High-resolution magnetization and/or transport measurements on the single-crystalline samples \cite{Feng_PRB_13} or the magnetic field-dependent ND measurements \cite{Kimura_PRL_08} can be helpful to further understand the nature and origin of these multiple field-induced magnetic phases in TbSi and TbSi$_{0.6}$Ge$_{0.4}$.

\subsection{Low-temperature crystallography}

\begin{figure*}
\centering
\includegraphics[width=1\textwidth]{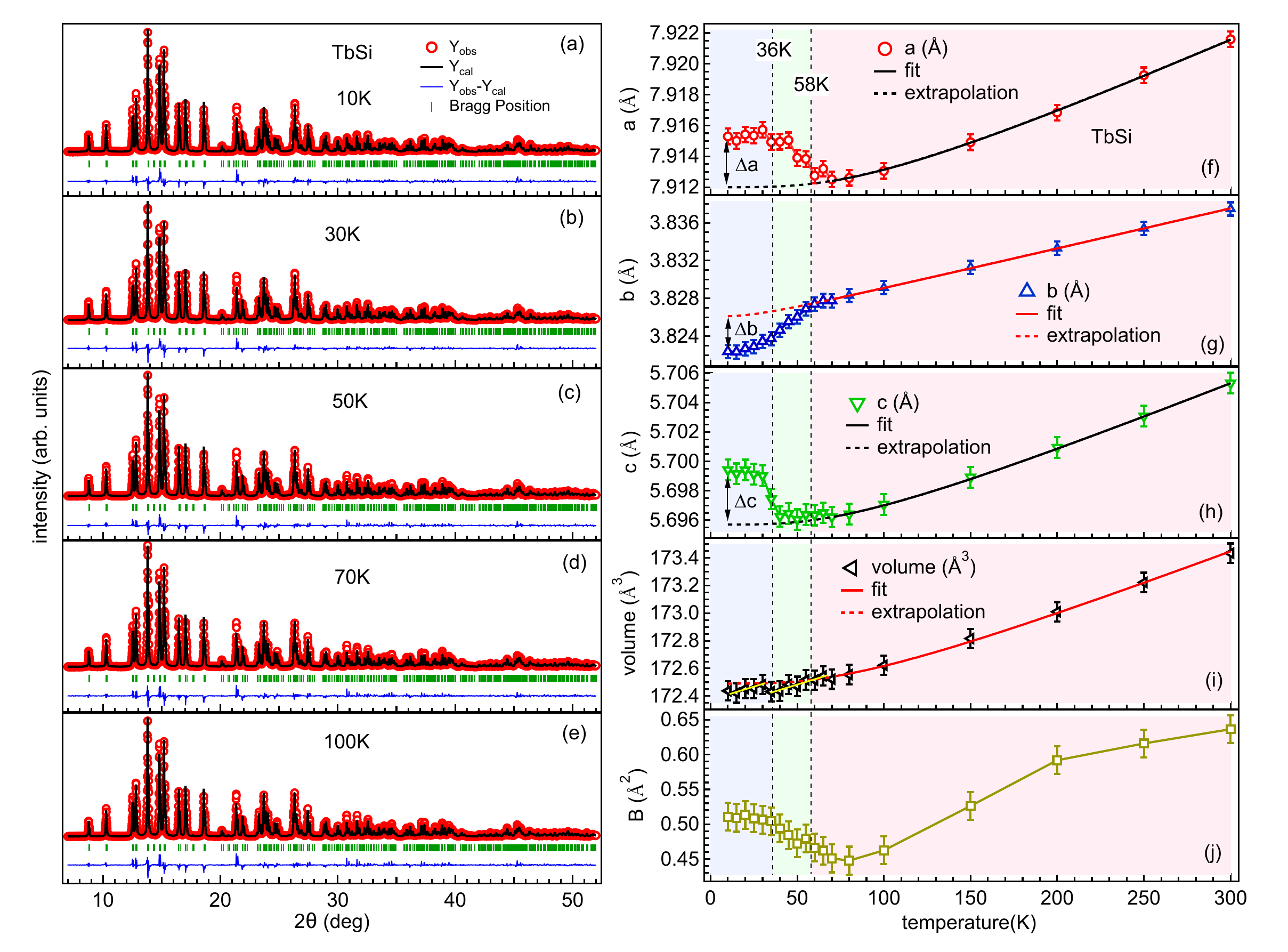}
\caption {(a--e) The Rietveld refinement of the XRD data of TbSi recorded using the Mo K$\alpha$ radiation at selected temperatures. The temperature dependence of the lattice parameters (f) $a$, (g) $b$, (h) $c$, (i) unit-cell volume, and (h) overall $B$ parameter, extracted from the Rietveld refinement. The vertical dashed lines represent the AFM transitions at 58 and 36~K, taken from the heat capacity measurements (Table I). The solid curves in (f--i) represent the best fit of the unit cell parameters from 70--300~K using the  Gr\"uneisen approximation up to first order and the dashed curves represent the extrapolation of the fitted data down to 10~K. The yellow lines in (i) are the guide to eye to show the discontinuity in cell volume at T$_{\rm N2}$.} 
\label{Fig13_LTXRD_TbSi}
\end{figure*}

The temperature-dependent XRD measurements have been performed on both samples from 10 to 300~K to understand the correlation between the crystal structure and magnetic properties of these compounds and further uncover the nature of their magnetic transitions. The selected Rietveld refinements of the powder XRD data of the TbSi sample recorded using the Mo K$_\alpha$ radiation are shown in Figs. \ref{Fig13_LTXRD_TbSi}(a--e). The refinement shows that the crystal structure of TbSi remains FeB-type (orthorhombic symmetry, $Pnma$ space group) down to 10~K. However, we observe a significant deviation in the lattice parameters near both magnetic transition temperatures, as shown in Figs. \ref{Fig13_LTXRD_TbSi}(f--j). This indicates the presence of strong magnetostriction in this sample. Note that a large temperature step of 5~K across the magnetic transitions unable us to distinguish the structural changes associated with the LT and IT AFM phases. The lattice parameters $a$ and $b$ show a notable elongation and contraction below T$_{\rm N1}$, as shown in Figs. \ref{Fig13_LTXRD_TbSi}(f, g), respectively. Interestingly, the $c$ parameter on the other hand shows no significant deviation from the conventional temperature-induced cell contraction (phonons) at T$_{\rm N1}$. In the HT AFM region of TbSi, the Tb moments lie in the $ab$ plane at an angle of 80$^o$ with the $a$-axis associated with wave vector q$\approx$(0, 1/2, 1/8) \cite{Papamantellos_JMMM_88} and hence change in only $a$ and $b$ lattice parameters is in accordance with the strong spin-lattice coupling in this sample. The lattice parameters $a$ and $c$ show a change in their behavior below T$_{\rm N2}$/T$_{\rm N3}$, where the lattice parameter $a$ remains almost invariant; however, the parameter $c$ shows an abrupt jump (elongation) below this temperature. The ND measurements demonstrate the first-order transformation from a commensurate magnetic structure with q = (0, 1/2, 0) to a three-dimensionally modulated phase having q = (q$_x$, q$_y$, q$_z$), with a most remarkable change only in the q$_z$ value from 0 to 0.126 between 36-39~K \cite{Papamantellos_JMMM_88}. Therefore, the observed first-order-like abrupt change in the $c$ lattice parameter in this temperate regime of TbSi is associated with the drastic modulation in the magnetic wave vector in the $z$-direction, indicating the sizable magnetostriction in the sample also at the order-order transition. To quantify the observed spontaneous magnetostriction in TbSi, the lattice parameters above the magnetic transition temperatures are fitted using the Gr\"uneisen function, given as \cite{Zhu_PRB_20, Kumar_JPCL_22}.

\begin{figure*}
\centering
\includegraphics[width=1\textwidth]{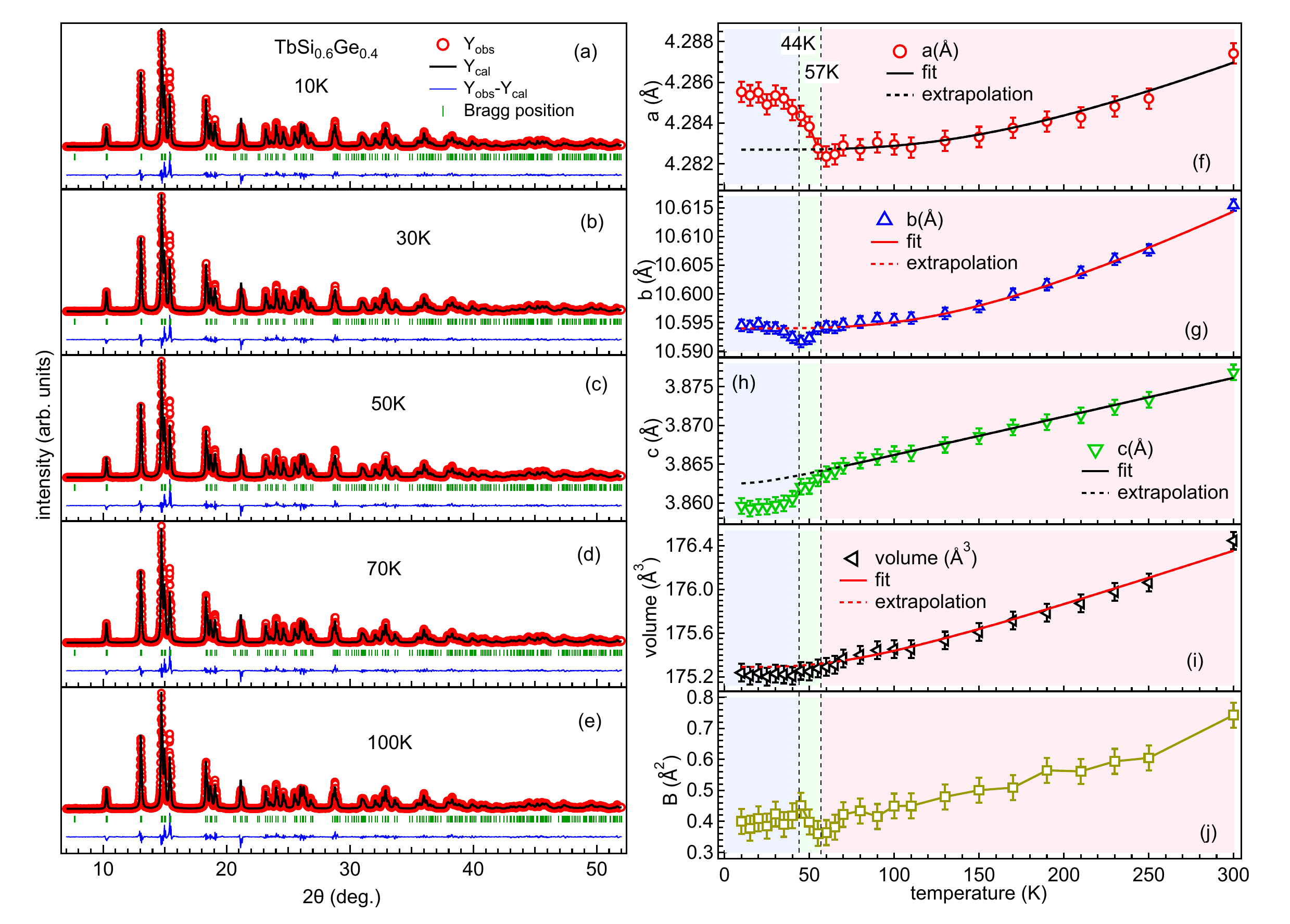}
\caption {(a--e) The Rietveld refinement of the powder XRD data of TbSi$_{0.6}$Ge$_{0.4}$ shown at some representative temperatures. The temperature dependence of the lattice parameters (f) $a$, (g) $b$, (h) $c$, (i) cell volume, and (j) overall $B$ parameter. The vertical dashed lines represent the AFM transitions at 57 and 44~K, taken from the heat capacity measurements (Table I). The solid curves in (f--i) represent the best fit of the unit cell parameters from 70--300~K using the  Gr\"uneisen approximation and the dashed curves represent the extrapolation of the fitted data down to 10~K. } 
\label{Fig14_LTXRD_TbSiGe}
\end{figure*}

\begin{equation} 
\epsilon(T)=\epsilon_0 + K_0U(T), 
\label{Gur1}
\end{equation} 

 where $\epsilon_0$ is a lattice constant value at 0~K, $K_0$ is the measure of the incompressibility of the sample (a constant), and $U(T)$ is internal energy, which can be expressed using the Debye function as 

\begin{equation}
U(T) = 9Nk_BT\left(\frac{T}{\theta_D}\right)^3 \int_0^{\theta_D/T} \frac{x^3}{(e^x-1)}dx, 
\label{Gur2}
\end{equation} 

where $N$ represents the effective number of atoms in a unit cell (eight in the present case) and $k_B$ is the Boltzmann constant. The solid curves in Figs. \ref{Fig13_LTXRD_TbSi}(f--i) represent the best fit of the lattice parameters from 70 to 300~K using the above equations, and the fitting parameters are given in Table III. Interestingly, the $b$ parameter shows a remarkably different temperature evolution not only in the magnetically ordered state below T$_{\rm N1}$ (discussed above) but also in the paramagnetic state. The lattice parameter $b$ shows almost linear temperature dependence between 70 and 300~K, whereas $a$ and $c$ parameters show significant curvature in this temperature range [see Figs. \ref{Fig13_LTXRD_TbSi}(f--h)], which is evident from the significantly lower value of the Debye temperature, $\theta_D$, for the $b$-axis as compared to $a$ and $c$, given in Table III. This highly anisotropic temperature dependence of the lattice parameters even in the PM state of TbSi may arise from the presence of large orbital angular momentum (L=3) and hence finite single-ion anisotropy of Tb$^{3+}$ ions \cite{Zou_PRB_07, Zou_PRB_08}. The calculated deviations in the lattice parameters at 10~K from the Gr\"uneisen function along $a$, $b$, and $c$ directions are 0.04\%, 0.1\%, and 0.06\%, respectively, as shown by the arrows in Figs. \ref{Fig13_LTXRD_TbSi}(f--h), where the deviation $\Delta$a$_{10K}$=100*(a$_{\rm exp}$-a$_{\rm fit}$)/a$_{\rm exp}$. Note that despite the first-order-like discontinuous change in the $c$ parameter, the overall relative deviation is higher in the case of $b$. The unit cell volume shows a small but notable change at T$_{\rm N2}$/T$_{\rm N3}$, as shown by a discontinuity in the yellow lines in Fig. \ref{Fig13_LTXRD_TbSi}(i). The overall Debye-Waller factor ($B$) increases below T$_{\rm N1}$, as shown in Fig. \ref{Fig13_LTXRD_TbSi}(j), indicating strong spin-lattice coupling in the sample. However, the $B$ parameter starts increasing from $\sim$80~K, i.e., well above the HT AFM region, which suggests the presence of short-range magnetic correlations and their coupling with the lattice even in the PM region.\par

\begin{table*}
\label{Table_fit}
\caption{Fit parameters extracted from the Gr\"uneisen approximation of the lattice parameters of TbSi and TbSi$_{0.6}$Ge$_{0.4}$ samples from 70 to 300~K.} 

\begin{tabular}{p{4cm}p{3cm}p{3cm}p{3cm}p{3cm}}
\hline
\hline
& a & b & c & V \\
\hline
& & TbSi & &\\
\hline
$\epsilon_0$ [\AA(\AA$^3$)] &7.9120(1)&3.8261(6) & 5.6957(3) & 172.49(2)\\
$K_0$ [\AA(\AA$^3$)/J] &1.53 x 10$^{17}$&1.29 x 10$^{17}$ &1.45 x 10$^{17}$ &1.47 x 10$^{19}$ \\	
$\theta_D$ (K)&350(7)&90(8) & 310(5) & 320(6)\\	
\hline
& & TbSi$_{0.6}$Ge$_{0.4}$ & &\\
\hline
$\epsilon_0$ [\AA(\AA$^3$)] &4.2827(3)&10.5940(2) & 3.8625(4) & 175.29(2)\\
$K_0$ [\AA(\AA$^3$)/J] &1.15 x 10$^{17}$ & 4.92 x 10$^{17}$ &1.52 x 10$^{17}$ &1.59 x 10$^{19}$ \\	
$\theta_D$ (K)& 715 (20) & 640(17) & 80(13) & 300(6)\\	         
\hline
\hline
\end{tabular}
\end{table*}

\begin{figure*}
\centering
\includegraphics[width=0.9\textwidth]{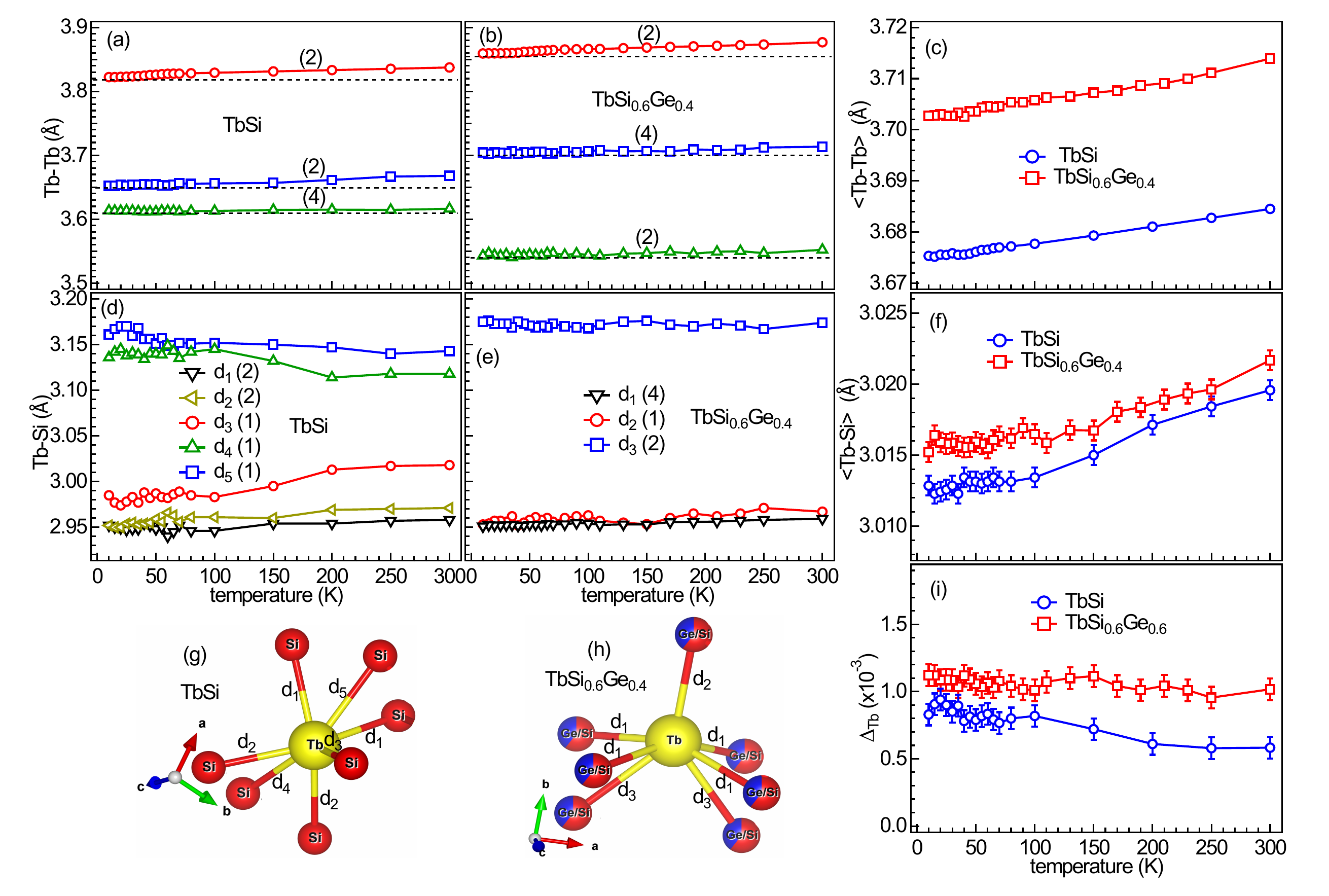}
\caption {(a, b) The temperature dependent first, second, and third nearest neighbour Tb-Tb distances for TbSi and TbSi$_{0.6}$Ge$_{0.4}$ samples, respectively. The number in the parenthesis represent the degeneracy of the respective interatomic distances. The horizontal dashed lines are drawn only to show the temperature variation in these Tb-Tb distances. (c) Average Tb-Tb distances as a function of temperature for both the samples. (d, e) Temperature dependence of neighbouring Tb-Si/Ge distances for TbSi and TbSi$_{0.6}$Ge$_{0.4}$ samples, respectively. (f) Average Tb-Si/Ge distances for both the samples. (g, h) A 3D representation of the Tb atoms surrounded by seven Si/Ge atoms in TbSi and TbSi$_{0.6}$Ge$_{0.4}$ crystals, respectively. (i) Temperature dependent local distortion parameter ($\Delta$) around Tb atoms for both the samples (see text for more details).} 
\label{Fig15_LTXRD_local}
\end{figure*}

In Figs. \ref{Fig14_LTXRD_TbSiGe}(a-e), we show the refined powder XRD patterns of TbSi$_{0.6}$Ge$_{0.4}$ at the same selected temperatures as for TbSi. The crystal structure of TbSi$_{0.6}$Ge$_{0.4}$ also remains the same, viz., CrB-type (orthorhombic symmetry, $Cmcm$ space group), in the whole studied temperature range down to 10~K. The temperature dependence of the Rietveld refined lattice parameter of TbSi$_{0.6}$Ge$_{0.4}$ is shown in Figs. \ref{Fig14_LTXRD_TbSiGe}(f--j). The lattice parameters $a$ and $c$, shown in Figs. \ref{Fig14_LTXRD_TbSiGe}(g, i), respectively, exhibit elongation and contraction below T$_{\rm N1}$. Notably, below T$_{\rm N2}$, this effect slightly decreases along the $a$-axis but increases along the $c$-axis. The magnetic moments in the HT AFM phase of TbSi$_{0.6}$Ge$_{0.4}$ lie along the $c$-axis as well as in the $ac$ plane in equal proportion \cite{Papamantellos_JMMM_88}. Therefore, a change in the $a$ and $c$ parameters is attributed to the magnetostriction due to spin-lattice coupling in the sample. However, it is important to note that the lattice parameter $c$ shows higher magnetostriction in the LT AFM phase, despite the fact that all the Tb moments lie only in the $ac$-plane in the LT phase of TbSi$_{0.6}$Ge$_{0.4}$ \cite{Papamantellos_JMMM_88}. This shows that the planar AFM with q=(1/2, 0, 1/2) in the LT phase produces more (less) magnetostriction along the $c$ ($a$)-axis compared to the sinusoidally modulated AFM having q=(q$_x$, 0, q$_z$) in the HT AFM phase region of TbSi$_{0.6}$Ge$_{0.4}$. More strikingly, the lattice parameter $b$ in TbSi$_{0.6}$Ge$_{0.4}$ first shows contraction below T$_{\rm N1}$ and then elongation below T$_{\rm N2}$, despite having no magnetic moments lying along the $b$-axis in any of the magnetically ordered phases. This indicates that any change in the $b$-parameter in the two AFM regions is a consequence of change in the unit cell along the $a$ and $c$-directions due to finite magnetoelastic coupling in the sample. The fitting of high-temperature lattice parameters of TbSi$_{0.6}$Ge$_{0.4}$ from 70 to 300~K using equations \ref{Gur1} and \ref{Gur2} [shown by solid lines in Figs. \ref{Fig14_LTXRD_TbSiGe}(f--i)] gives significantly lower (around one order of magnitude) Debye temperature for the $c$ parameter compared to $a$ and $b$ (see Table III). Note that the lattice parameter $b$ in $Pnma$ setting is equivalent to the $c$ parameter in the $Cmcm$ symmetry \cite{Papamantellos_JLCM_85, Jiang_NC_23}. This explains the similar temperature-dependent behavior of the lattice parameter $c$ in TbSi$_{0.6}$Ge$_{0.4}$ to the $b$ parameter in TbSi and confirms the highly anisotropic thermal expansion of the lattice parameters in both the samples in the magnetically ordered as well as disordered (PM) states. Further, the unit cell volume of TbSi$_{0.6}$Ge$_{0.4}$, shown in Fig. \ref{Fig14_LTXRD_TbSiGe}(i), increases compared to the TbSi sample in accordance with the larger atomic size of the Ge atoms than Si \cite{Shannon_AC_76}. However, interestingly, the unit cell volume of TbSi$_{0.6}$Ge$_{0.4}$ displays no significant change at T$_{\rm N2}$, making it an interesting case where the deviation in the unit cell volume associated with the first-order magnetic transition is negligible. The compounds exhibiting an abrupt change in magnetization at the first-order transition accompanied by a small change in the cell volume are desirable for magnetic refrigeration applications owing to their large value of $\Delta S_{\rm M}$ along with better cycle stability. Recently, RE$_2$In (RE=Eu and Pr) compounds were also found to display a similar behavior, with a volume change of as low as $\sim$0.1\% at their first-order FM$\rightarrow$PM transition \cite{Guillou_NC_18, Biswas_JSSST_22, Biswas_PRB_20}. The absence of any discontinuous change in the cell volume along with a significant broadening in the magnetic transition at the higher field and negligible thermal hysteresis make TbSi$_{0.6}$Ge$_{0.4}$ a possible candidate to use for magnetocaloric applications. While in TbSi$_{0.6}$Ge$_{0.4}$ the change in magnetic entropy is moderate, our work indicates that such almost anhysteretic first-order transitions associated with small volume changes may be abundant in rare earth compounds. We observed a change in the overall Debye-Waller factor at both T$_{\rm N1}$ and T$_{\rm N2}$ in the case of TbSi$_{0.6}$Ge$_{0.4}$ as shown in Fig. \ref{Fig14_LTXRD_TbSiGe}(j), keeping the large error associated with the B-factor in the Rietveld refinement in mind.\par

Now we discuss the change in the local coordination environment of these compounds with Ge substitution. In Figs. \ref{Fig15_LTXRD_local}(a, b), the temperature dependence of the first, second, and third nearest neighbor (NN) Tb-Tb distances is plotted for TbSi and TbSi$_{0.6}$Ge$_{0.4}$, respectively, where the numbers in parentheses show their respective degeneracies. Each Tb atom is coordinated with eight different Tb atoms with varying distances in both $Pnma$ and $Cmcm$ structures. Interestingly, in the case of TbSi, the first NN Tb-Tb distance is four-fold degenerate and the second NN distance is two-fold degenerate, whereas the opposite trend is observed in TbSi$_{0.6}$Ge$_{0.4}$ [see Figs. \ref{Fig15_LTXRD_local}(a, b)]. The third NN Tb-Tb bond distance is doubly degenerate in both cases. In fact, this third NN Tb-Tb bond distance represents the unit cell parameters $b$ and $c$, respectively, for TbSi and TbSi$_{0.6}$Ge$_{0.4}$ samples in their respective crystal structures \cite{Papamantellos_JLCM_85, Jiang_NC_23}. The increase in the interatomic distance between the Tb ions for both samples occurs on heating in accordance with general lattice expansion [the horizontal dashed lines in Figs. \ref{Fig15_LTXRD_local}(a, b) are the baselines]. The temperature-dependent average Tb-Tb distances are plotted in Fig. \ref{Fig15_LTXRD_local}(c) for both compounds. An increase in the average Tb-Tb distances has been observed with temperature and Ge substitution due to thermal and chemical expansion of the unit cell, respectively. Each Tb atom is surrounded by seven Si/Ge atoms in both samples. We note that the $Pnma$ space group has lower crystal symmetry compared to $Cmcm$. Therefore, in the case of TbSi, we observe a total of five different Tb-Si bond distances, with two sets of two equal (1$^{\rm st}$ and 2$^{\rm nd}$ NN) and three unequal (3$^{\rm rd}$, 4$^{\rm th}$, and 5$^{\rm th}$ NN) bonds, as schematically represented in Fig. \ref{Fig15_LTXRD_local}(g). Whereas, only three non-equivalent Tb-Si/Ge bond distances were observed in TbSi$_{0.6}$Ge$_{0.4}$ with degeneracies of four, one, and two for the first, second, and third NN, respectively, as represented in Fig. \ref{Fig15_LTXRD_local}(h). The temperature evolution of these Tb-Si/Ge bonds for TbSi and TbSi$_{0.6}$Ge$_{0.4}$ is presented in Figs. \ref{Fig15_LTXRD_local}(d) and (e), respectively. Here, a number in the subscript of letter d represents the order of the NN distance of the Tb atom, from the shortest to the longest. Due to increasing lattice symmetry, d$_1$ and d$_2$ of TbSi merge and become d$_1$ in the case of TbSi$_{0.6}$Ge$_{0.4}$, whereas d$_4$ and d$_5$ become d$_3$. The temperature evolution of the average Tb-Si/Ge distance is presented in Fig. \ref{Fig15_LTXRD_local}(f) for both samples, showing the same increase with temperature and Ge doping as the Tb-Tb distances. In order to further understand the change in the local coordination environment around the Tb atoms, we calculate the local distortion parameter, $\Delta$, defined as \cite{Zhu_PRB_20, Kumar_PRB_22}.

\begin{equation}
 \Delta = \frac{1}{n} \sum_{n=1}^n \left( \frac{d_n-<d>}{<d>}\right)^2, 
\label{distortion}
\end{equation} 

where n represents the number of Si/Ge atoms coordinated to Tb (seven in this case), d$_n$ represents each Tb-Si/Ge bond distance, and $<d>$ is the average Tb-Si/Ge distance. The temperature-dependent local distortion parameter around Tb atoms (by Si/Ge atoms only) is presented in Fig. \ref{Fig15_LTXRD_local}(i) for both samples. It is interesting to note that the distortion parameter, $\Delta$, increases in TbSi$_{0.6}$Ge$_{0.4}$  compared to the TbSi sample, despite the higher overall crystal symmetry of the former. Thus, the change from the $Pnma$ structure in our system towards the $Cmcm$ is coupled with the enhancement in the local distortion within the unit cell. Interestingly, one may argue that the bond distances obtained by Rietveld refinement are already averaged with respect to Si/Ge distribution, and further averaging of atomic distances should blunt all the details about the local structure. To the contrary, the $\Delta$ parameter accurately identifies the possibility of lattice changes due to distortion caused by chemical substitution, making it useful for future predictions.

\section{\noindent ~Conclusion}

In summary, theoretical investigations reveal that a minimal 2.5\% Ge doping is sufficient to induce a structural transition from the FeB-type $Pnma$ lattice to the CrB-type $Cmcm$ in TbSi$_{1-x}$Ge$_x$ ($x =$0--1) compounds. This transition is attributed to the increased local distortion resulting from Ge substitution, favoring the emergence of the higher symmetry $Cmcm$ structure. Subsequent experimental analyses, encompassing magnetization, specific heat, and temperature-dependent X-ray diffraction (XRD), are conducted on TbSi and TbSi$_{0.6}$Ge$_{0.4}$ samples. TbSi crystallizes in the FeB-type crystal structure ($Pnma$ space group), while TbSi$_{0.6}$Ge$_{0.4}$ adopts the CrB-type structure ($Cmcm$ space group). Both TbSi and TbSi$_{0.6}$Ge$_{0.4}$ samples exhibit two magnetic transitions: a high-temperature (HT) antiferromagnetic (AFM) to paramagnetic (PM) transition at T$_{\rm N1}\approx$ 58 K and 57 K, and a low-temperature (LT) AFM to HT AFM transition at T$_{\rm N2}\approx$ 36 K and 44 K, respectively. Detailed C$_{\rm P}$ measurements manifest the second- and first-order nature of these phase transitions, respectively, in both samples. We unravel an additional intermediate temperature (IT) AFM phase between T$_{\rm N1}$ and T$_{\rm N2}$ in the case of TbSi, where both LT to IT and IT to HT AFM phase transitions exhibit first-order nature. Both samples display a metamagnetic-like behavior with a critical magnetic field of around 6 T at 50 K, indicative of weak HT AFM interactions under an external magnetic field. The TbSi and TbSi$_{0.6}$Ge$_{0.4}$ samples show a magnetic entropy change ($\Delta S_{\rm M}$) of 9.6 (and $\Delta$T$_{\rm ad}\approx$ 1.2 K) and 11.6 J/kg-K, respectively, at T$_{\rm N2}$ for $\Delta\mu_0$H= 7 T. The TbSi$_{0.6}$Ge$_{0.4}$ sample displays a significant broadening in the $\Delta$S$_{\rm M}$ curves at higher magnetic fields, yielding a reasonably large value of TEC(10)=9.3 J/kg-K as compared to TbSi (5.4 J/kg-K) at $\Delta \mu_0$H=7 T. The TbSi$_{0.6}$Ge$_{0.4}$ and, to a lesser extent, TbSi display field-induced additional AFM interactions in the HT phase region, which are found to be responsible for the significant broadening in the $\Delta S_{\rm M}$ curves of the former at higher magnetic fields. Moreover, both TbSi and TbSi$_{0.6}$Ge$_{0.4}$ exhibit highly anisotropic thermal expansion behavior in both AFM and PM states, accompanied by notable magnetostriction and magnetoelastic coupling at the observed transition temperatures. Finally, a detailed H-T phase diagram for both samples has been constructed utilizing the C$_{\rm P}$(H, T) data, elucidating their complex magnetic behavior under varying temperature and magnetic fields.

\subsection{Acknowledgments}

This work was supported by the U.S. Department of Energy (DOE), Office of Science, Basic Energy Sciences, Materials Science and Engineering Division. The research was performed at Ames National Laboratory, which is operated for the U.S. DOE by Iowa State University under contract \# DE-AC02-07CH11358.

\end{document}